\documentclass[12pt]{article}
\usepackage{epsfig}
\newcommand{\mysection}{\setcounter{equation}{0}\section}

\def\beq{\begin{equation}}
\def\eeq{\end{equation}}
\def\beqa{\begin{eqnarray}}
\def\eeqa{\end{eqnarray}}

\newlength{\dinwidth} \newlength{\dinmargin}
\begin{document}
\begin {flushright}
FSU-HEP-20000920\\
\end {flushright} 
\vspace{3mm}
\begin{center}
{\Large \bf High-order corrections and subleading logarithms
for top quark production}
\end{center}
\vspace{2mm}
\begin{center}
{\large Nikolaos Kidonakis}\\
\vspace{2mm}
{\it Physics Department\\
Florida State University\\
Tallahassee, FL 32306-4350, USA} \\
\end{center}

\begin{abstract}
We derive high-order threshold corrections for 
top quark production in hadronic collisions
from resummation calculations. 
We present analytical expressions for the cross section through 
next-to-next-to-next-to-next-to-leading order (N$^4$LO)
and next-to-next-to-leading logarithmic accuracy. 
Special attention is paid to the role of subleading logarithms
and how they relate to the convergence of the perturbation series
and differences between various resummation prescriptions. 
It is shown that care must be taken to avoid unphysical terms
in the expansions.
Numerical results are presented for top quark production at the Tevatron.
We find sizeable increases to the total cross section
and differential distributions and a dramatic reduction of the 
factorization scale dependence relative to next-to-leading order.
\end{abstract}
\pagebreak

\mysection{Introduction}

The top quark production cross section at the Tevatron receives
significant contributions from the threshold region, where there is
limited phase space for the emission of real gluons. The 
incomplete cancellation of infrared divergences between real and
virtual graphs produces finite, but large, logarithmic corrections
in the form of ``plus'' distributions. These corrections can be resummed
to all orders in the perturbative expansion.

The need for threshold resummations was recognized over a decade ago
for the Drell-Yan cross section \cite{Sterman:1987aj,Catani:1989ne}. 
The resummation of the leading logarithms for heavy quark production
\cite{Laenen:1992af,Berger:1996ad,Catani:1996yz} 
relies heavily on this work since these logarithms are universal
between electroweak and QCD induced cross sections. To go beyond leading 
logarithms one has to take into account the complex color structures 
of QCD cross sections \cite{Kidonakis:1996aq,Kidonakis:1997gm}. 
Resummation for heavy quark production at next-to-leading logarithmic (NLL) 
accuracy was first presented in Refs. \cite{Kidonakis:1996aq,Kidonakis:1997gm}
and then in Refs. \cite{Kidonakis:1997zd,
Kidonakis:1997ed,Bonciani:1998vc,Laenen:1998qw,Kidonakis:1998ei}.
For a review see Refs. \cite{Kidonakis:2000ze,Kidonakis:1998qe}.

Threshold resummation follows from the factorization properties
of the cross section \cite{Contopanagos:1997nh}, 
separating long- and short-distance physics.
The hadronic cross section is written as a convolution 
of non-perturbative parton distribution functions with the perturbative
partonic cross section.
This convolution becomes a simple product if one takes 
moments of the cross section.
The perturbative cross section still has sensitivity to soft-gluon
dynamics and is a smooth function only away from the
edges of partonic phase space. In general it includes ``plus''
distributions with respect to a variable that measures distance
from partonic threshold. It is these singular distributions
that can be resummed to all orders in perturbation theory. This
is achieved by first refactorizing the cross section \cite{Kidonakis:1997gm} 
into new functions which absorb the universal collinear 
singularities associated with the incoming partons, and a function 
that describes non-collinear soft gluon emission.
Resummation is explicitly derived in moment space from the 
renormalization group properties of these functions 
\cite{Kidonakis:1997gm,Laenen:1998qw,Kidonakis:1998bk,Kidonakis:1998nf}.
To obtain the physical resummed cross section the moment space
results must be inverted back to momentum space. There have been a few
proposals or ``prescriptions'' for the best way to do the inversion 
or otherwise use the resummation 
\cite{Laenen:1992af,Berger:1996ad,Catani:1996yz}.
Numerically, the choice of prescription
can have a significant effect. 
Alternatively, the resummed cross section can be used as a
generator of perturbation theory. At fixed order no prescription
is necessary to invert the moment-space results. This is the approach taken
in this paper.

The aim of this paper is two-fold. 
First, to derive the expansion of the resummed cross section through 
next-to-next-to-next-to-next-to-leading order (N$^4$LO) 
and next-to-next-to-leading logarithmic (NNLL) accuracy; and second, 
to investigate and assess the importance of 
subleading logarithms (beyond NNLL). 
We calculate the numerical effect of these higher orders and
subleading logarithms for top quark production at the Tevatron.
As we will see, this discussion intimately relates
to the convergence of the perturbation series as well as 
to the choice of a resummation prescription, over which there have been 
different viewpoints \cite{Catani:1996yz,Berger:1998gz}.
We will show that one has to be careful to avoid unphysical terms
in the fixed-order expansions.
These considerations apply to QCD hard scattering cross sections in general.
We find large corrections for top quark production at the Tevatron
and a dramatically reduced factorization scale dependence.
Our formalism allows us to make predictions for both total and differential
cross sections.

In a companion paper \cite{KLMV} the next-to-leading order (NLO) 
and next-to-next-to-leading order (NNLO) 
threshold corrections to heavy quark production at NNLL accuracy are 
also studied and the difference
between single-particle inclusive (1PI) and pair inclusive (PIM) kinematics 
is highlighted.
Similar NNLO expansions have recently been presented for 
electroweak-boson \cite{Kidonakis:2000ur,Kidonakis:1999xm}, 
direct photon \cite{Kidonakis:2000hq,Kidonakis:1999mk}, 
and jet \cite{Kidonakis:2001gi} hadroproduction.

In the next section we briefly review the resummation formalism
and give the expression for the resummed cross section in moment space.
In Section 3 we present NLO and NNLO expansions at NNLL accuracy
in single-particle inclusive kinematics and we discuss 
in detail the role of subleading logarithms. 
We also present numerical results for the top
total cross section and transverse momentum distributions at the Tevatron. 
In Section 4 we present the next-to-next-to-next-to-leading order (N$^3$LO) 
corrections at NNLL accuracy, and again we study the role of 
subleading logarithms at that order. In Section 5 we discuss
N$^4$LO and higher-order corrections. 
In Appendix A we list several formulas
for the Mellin transforms that are used in the calculations. 
In Appendix B we collect some formulas for the integrations involved in 
the calculation of the hadronic total and differential cross sections.
In Appendix C we present results for the NLO and NNLO expansions in
PIM kinematics.

\mysection{Resummed cross section}

We begin by briefly reviewing the resummation formalism of
Refs. \cite{Kidonakis:1996aq,Kidonakis:1997gm,Laenen:1998qw}.
We will mostly discuss the single-particle inclusive cross section 
but will make a few comments on pair inclusive kinematics where appropriate.
The differences between the two kinematics in the expression for
the resummed cross section are minimal. 

The factorized single-heavy quark inclusive cross section 
for hadron-hadron collisions,
\beq
h_A(p_A)+h_B(p_B) \rightarrow Q(p_1) + X[\bar Q] \, ,
\eeq
where the $h$'s are the colliding hadrons, 
$Q$ is the produced heavy quark, and $X$ represents the additional
partons in the final state including the heavy antiquark,
takes the form of a convolution of the perturbative short-distance
cross section ${\hat \sigma}$ with the universal parton distribution 
functions $\phi$:
\beq
\sigma_{h_Ah_B\rightarrow QX}=\int dx_a \, dx_b \; \phi_{f_a/h_A}(x_a,\mu_F^2)
\; \phi_{f_b/h_B}(x_b,\mu_F^2) \; 
{\hat \sigma}_{f_af_b\rightarrow QX} \left(s_4,t_1,u_1,m^2,\mu_F^2,
\alpha_s(\mu_R^2)\right) \, ,
\label{factcs}
\eeq
where $\mu_F$ and $\mu_R$ are the factorization and renormalization
scales, respectively.
Note that here and in the following $\sigma$ and ${\hat \sigma}$ 
can denote either the total cross section or any relevant differential
cross section. 
The parton processes involved at the Born level are
\beqa
q(p_a)+{\bar q}(p_b) &\rightarrow& Q(p_1) + {\bar Q}(p_2) \, ,
\nonumber \\ 
g(p_a)+g(p_b) &\rightarrow& Q(p_1) + {\bar Q}(p_2) \, .
\eeqa
The $q{\bar q}$ channel is dominant for top production
at the Tevatron and contributes over 90\% of the cross section 
at the Born level.
The partonic invariants in Eq. (\ref{factcs}) are defined by
\beq
s=(p_a+p_b)^2, \quad t_1=(p_a-p_1)^2-m^2, 
\quad  u_1=(p_b-p_1)^2-m^2 \, ,
\eeq
with $m$ the heavy quark mass,
while $s_4=s+t_1+u_1$ measures the distance from threshold;
at threshold $s_4=0$. 

By taking moments and replacing the incoming hadrons by partons, 
the convolution in Eq. (\ref{factcs}) simplifies to a product
\cite{Kidonakis:1997gm,Laenen:1998qw}, 
\beq
{\tilde{\sigma}}_{f_a f_b \rightarrow QX}(N)
={\tilde{\phi}}_{f_a/f_a}(N_a)\,  {\tilde{\phi}}_{f_b/f_b}(N_b)\,
\hat{\sigma}_{f_a f_b \rightarrow QX}(N) \, .
\label{fact}
\eeq
The moments are defined by 
$\hat{\sigma}(N)=\int (ds_4/s) \, 
e^{-Ns_4/s} {\hat\sigma}(s_4)$, with $N$ the moment variable,
and similarly for the $\phi_i$'s with respect to $x_i$.
The definition of $N_i$ depends on the kinematics.
For 1PI kinematics, $N_a=N(-u/s)$ and $N_b=N(-t/s)$ \cite{Laenen:1998qw}.
For PIM kinematics, $N_a=N_b=N$ \cite{Kidonakis:1997gm}.

The short-distance perturbative cross section $\hat {\sigma}$ 
still displays sensitivity to soft gluon
emission. The incomplete cancellation between graphs with gluon
emission and virtual gluon corrections manifests itself in
the occurence in $\hat {\sigma}$ of ``plus'' distributions which are 
singular at $s_4=0$, the partonic threshold.
At $n$th order in $\alpha_s$ (beyond the Born term), 
these distributions are of the form
$[(\ln^k(s_4/m^2))/s_4]_+$, $k \le 2n-1$.
The leading logarithms correspond to $k=2n-1$, NLL to $k=2n-2$,
NNLL to $k=2n-3$ and so on.
Under moments $[(\ln^{2n-1}(s_4/m^2))/s_4]_+ \rightarrow \ln^{2n}N$ (see 
Appendix A), and our goal becomes to resum logarithms of $N$.

To separate these soft gluon effects from the hard scattering,
a refactorization is introduced 
\cite{Kidonakis:1996aq,Kidonakis:1997gm,Laenen:1998qw,Kidonakis:1998bk},
\beq
{\tilde{\sigma}}_{f_a f_b \rightarrow QX}(N)
={\tilde{\psi}}_{f_a/f_a}(N_a)\,  {\tilde{\psi}}_{f_b/f_b}(N_b)\;
H_{IJ}^{f_af_b} \;  {\tilde{S}}_{JI}^{f_af_b}(m/(N\mu_F)) \, , 
\label{refact}
\eeq
where the $\psi$'s are center-of-mass parton distributions 
\cite{Sterman:1987aj} that absorb the
universal collinear singularities associated with the initial-state
partons, and $S$ is the soft-gluon function that describes non-collinear
soft gluon emission. The mass of the heavy quarks protects the final state
from collinear singularities. 
$H$ is the hard-scattering function and is  free of soft-gluon effects
and thus independent of $N$.
$H$ and $S$ are matrices in the space of color exchanges 
($I,J$ are color indices) and differ for each partonic process.

Using Eqs. (\ref{fact}) and (\ref{refact}) to solve for 
the perturbative cross section ${\hat{\sigma}}$, we then have
${\hat{\sigma}}(N)=({\tilde{\psi}}_{f_a/f_a}{\tilde{\psi}}_{f_b/f_b}
/({\tilde{\phi}}_{f_a/f_a}{\tilde{\phi}}_{f_b/f_b}))
\; {\rm Tr}\,[H{\tilde{S}}]$.
After resumming the $N$-dependence in $\psi/\phi$ and $S$
\cite{Kidonakis:1996aq,Kidonakis:1997gm,Laenen:1998qw}, we obtain the
resummed heavy quark cross section at NLL accuracy\footnote{Note that although
we formally have NLL accuracy in the resummed exponent, after
matching with the exact NLO cross section we can determine all NNLL
terms in the finite-order expansions.} in moment space:
\beqa
{\hat{\sigma}}_{f_a f_b \rightarrow QX}(N) &=&   
\exp\left[ E^{(f_a)}(N_a)+E^{(f_b)}(N_b)\right] \; 
\exp \left[2\int_{\mu_F}^{m} {d\mu' \over \mu'}
\left(\gamma_a(\alpha_s(\mu'^2))
+\gamma_b(\alpha_s(\mu'^2))\right)\right]
\nonumber\\ && \hspace{-20mm} \times \,
\exp\left[4\int_{\mu_R}^{m}\frac{d\mu'}{\mu'} 
\beta(\alpha_s(\mu'^2))\right] \; 
{\rm Tr} \left \{H^{f_af_b}\left(\alpha_s(\mu_R^2)\right) \right.
\nonumber\\ && \hspace{-37mm} \times \left.
\bar{P} \exp \left[\int_m^{m/N} {d\mu' \over \mu'} \;
(\Gamma_S^{f_af_b})^\dagger\left(\alpha_s(\mu'^2)\right)\right] \;
{\tilde S}^{f_af_b} \left(1,\alpha_s(m^2/N^2) \right) \; 
P \exp \left[\int_m^{m/N} {d\mu' \over \mu'}\; \Gamma_S^{f_af_b}
\left(\alpha_s(\mu'^2)\right)\right] \right\} \, .
\nonumber \\ 
\label{resHQ}
\eeqa
This expression is actually valid for both 1PI and PIM kinematics
with appropriate definitions for $N_a$ and $N_b$ as discussed previously.

The first exponent in Eq. (\ref{resHQ}) resums the $N$-dependence of the ratio
${\tilde \psi}_{f_i/f_i}/{\tilde \phi}_{f_i/f_i}$ and is given in 
the $\overline{\rm MS}$ scheme by 
\beq
E^{(f_i)}(N_i)=
-\int^1_0 dz \frac{z^{N_i-1}-1}{1-z}\;
\left \{\int^{\mu_F^2}_{(1-z)^2s} \frac{d\mu'^2}{\mu'^2}
A^{(f_i)}\left[\alpha_s\left({\mu'}^2\right)\right]
+\frac{1}{2}\kappa^{(f_i)}\left[\alpha_s((1-z)^2 s)\right]
\right\} \, .
\label{Eexp}
\eeq
At next-to-leading order accuracy in $\ln N$,
we need $A^{(f)}$ at two-loops,  
$A^{(f)}(\alpha_s) = C_f [ {\alpha_s/\pi}$ \linebreak
${}+({\alpha_s/\pi})^2 K/2 ]$, and
$\kappa^{(f)}=2C_f\; (\alpha_s/\pi) (1-\ln(2 \nu_f))$.
Here $C_f=C_F=(N_c^2-1)/(2N_c)$ for an incoming quark,
and $C_f=C_A=N_c$ for an incoming gluon, with $N_c$ the number of colors,
while $K= C_A\; ( 67/18-\pi^2/6 ) - 5n_f/9$,
where $n_f$ is the number of quark flavors.
The $\nu_i$ terms are gauge dependent. They are defined by
$\nu_i \equiv (\beta_i \cdot n)^2/|n|^2$,
where $\beta_i=p_i {\sqrt {2/s}}$ are the particle velocities
and $n$ is the axial gauge vector.
We note that all gauge dependence cancels out in the cross section.

In the DIS scheme, which is usually only applied to quarks,
the first exponent in Eq. (\ref{resHQ}) is given in terms of the 
$\overline{\rm MS}$ result as
\beq
E^{(q)}(N_i)|_{\rm{DIS}} =
E^{(q)}(N_i)|_{\rm{\overline{MS}}}
-\int_0^1 dz \frac{z^{N_i-1}-1}{1-z} \, 
\left\{\int^{1-z}_{1} \frac{d\lambda}{\lambda}
A^{(q)}\left[\alpha_s(\lambda s)\right]
+ B^{(q)} \left[\alpha_s((1-z)s)\right]\, \right\} \, ,
\eeq
where
$B^{(q)}(\alpha_s) = - (3C_F/4)(\alpha_s/\pi).$

The $\gamma_a$ are anomalous dimensions of the fields $\psi$ 
and are given at one loop by 
$\gamma_q=(\alpha_s/\pi)(3C_F/4)$ and $\gamma_g=(\alpha_s/\pi)(\beta_0/4)$ 
for quarks and gluons, respectively.
The $\beta$ function is given by
$\beta(\alpha_s) \equiv \mu \, d \ln g/d \mu
=-\beta_0 \alpha_s/(4 \pi) +...$, with $\beta_0=(11C_A-2n_f)/3$.

The trace appearing in the resummed expression is taken in color space.
The symbols $P$ and ${\bar P}$ denote path ordering in the same sense
as the variable $\mu'$ and against it, respectively.
The evolution of the soft function from scale $m/N$ to $m$ follows
from its renormalization group properties and 
is given in terms of the soft anomalous dimension matrix $\Gamma_S$
\cite{Kidonakis:1997gm}. 
For the determination of $\Gamma_S$ an appropriate choice of color basis 
has to be made.
For the $q{\bar q}$ channel we use an $s$-channel singlet-octet basis,
while for the $gg$ channel we use a basis consisting 
of three color tensors \cite{Kidonakis:1997gm}.
The soft anomalous dimension matrix, evaluated through the calculation of
one-loop eikonal vertex corrections, has been presented for the partonic
processes in heavy quark production 
in Refs. \cite{Kidonakis:1996aq,Kidonakis:1997gm,Kidonakis:2000ze}.
In the color bases that we use, the soft matrices, $S$, are diagonal for both
partonic channels, and the 
hard matrix for the $q{\bar q}$ channel has only one non-zero element.
At lowest order, the trace of the product of the hard and soft matrices
reproduces the Born cross section in each partonic channel.
We also note that the $\Gamma_S$ matrices are not diagonal in the color
bases that we use. If we perform a diagonalization so that
the $\Gamma_S$ matrices do become diagonal, then the path-ordered
exponentials of matrices in the resummed expression
reduce to simple exponentials; however, this diagonalization
procedure is complicated in practice \cite{Kidonakis:2000ze}.

The integrations over $z$ in the exponents of 
Eqs. (\ref{resHQ}), (\ref{Eexp}) run over the region
where the running coupling constant $\alpha_s$ diverges.
The prescriptions of Refs. \cite{Laenen:1992af,Berger:1996ad,Catani:1996yz} 
have been proposed to avoid these soft 
gluon divergences in the resummed cross section.
However, if we expand the exponents in the resummed cross section at
fixed order in $\alpha_s$ and invert back to momentum space using the
equations in Appendix A, no divergences are encountered and thus no
prescription is required.
In addition to avoiding the necessity for a resummation prescription, 
a finite-order expansion bypasses the need for the diagonalization procedure
that we mentioned above, as well. 

\mysection{NLO and NNLO threshold corrections}

In this section we expand the resummed cross section to next-to-leading
and next-to-next-to-leading orders.
In the following $\sigma^{(n)}$ stands for the $n$th-order 
differential corrections. Nominally, it denotes 
$s^2\, d^2\sigma^{(n)}/(dt_1du_1)$
but it can also denote any other relevant differential cross section,
such as $d^2\sigma^{(n)}/(dp_T^2 ds_4)$, with $p_T$ the transverse momentum, 
or $d^2\sigma^{(n)}/(dy ds_4)$, with $y$ the rapidity, or 
$d^2\sigma^{(n)}/(dp_T dy)$, with appropriate Jacobians inserted into the
definition of the Born term, $\sigma^B$,  and the function
$B_{\rm QED}$ for the $gg$ channel in the expressions below. 

\subsection{NLO threshold corrections} 

We first expand the resummed cross section to 
next-to-leading order in 1PI kinematics. 
These expansions are already known
for both 1PI and PIM kinematics \cite{Kidonakis:1997gm,Laenen:1998qw, 
Kidonakis:2000ze}. 

For the $q{\bar q}$ channel in the $\overline{\rm MS}$ scheme, 
the full next-to-leading-order threshold corrections are
\begin{eqnarray}
{\hat \sigma}^{\overline{\rm MS} \, (1)}_{q{\bar q}\rightarrow Q{\bar Q}}
(s_4,m^2,s,t_1,u_1,\mu_F,\mu_R)&=&\sigma^B_{q{\bar q}\rightarrow Q{\bar Q}}
\frac{\alpha_s(\mu_R^2)}{\pi}
\left\{4C_F\left[\frac{\ln(s_4/m^2)}{s_4}\right]_{+}\right.
\nonumber \\ && \hspace{-45mm} \left.
{}+\left[\frac{1}{s_4}\right]_{+} 
\left[2 {\rm Re} {\Gamma'}_{22}^{q{\bar q}}
-2C_F+2C_F\ln\left(\frac{sm^2}{t_1u_1}\right)
-2C_F\ln\left(\frac{\mu_F^2}{m^2}\right)\right]\right\}
\nonumber \\ && \hspace{-45mm} 
{}+\delta(s_4) \, \sigma^{(1)\, q {\bar q} \, 
{\rm S+V}}_{{\overline {\rm MS}}} \, ,
\label{NLOqq}
\end{eqnarray}
where  $\sigma^{(1)\,q {\bar q} \,{\rm S+V}}_{{\overline {\rm MS}}}$ 
denotes the soft plus virtual ($S+V$)
$\delta(s_4)$ terms in the NLO cross section that can be obtained 
from Eq. (4.7) in Ref. \cite{Beenakker:1991ma} (with $t_1$ and $u_1$ 
interchanged because of different definitions in that reference). 
Also
\beqa
{\rm Re}{\Gamma'}_{22}^{q{\bar q}}&=&C_F\left[4\ln\left(\frac{u_1}{t_1}\right)
-{\rm Re} L_{\beta}\right]
+\frac{C_A}{2}\left[-3\ln\left(\frac{u_1}{t_1}\right)
-\ln\left(\frac{m^2s}{t_1u_1}\right)
+{\rm Re} L_{\beta}\right]
\nonumber \\ &&
\eeqa
is obtained from the real part of the one-loop soft anomalous dimension matrix 
element $\Gamma_{22}^{q{\bar q}}$ after dropping all 
gauge dependent terms and an overall coefficient $\alpha_s/\pi$.
Here $L_{\beta}=(1-2m^2/s)/\beta\cdot[\ln((1-\beta)/(1+\beta))+\pi i]$,
with $\beta=\sqrt{1-4m^2/s}$,
is the velocity-dependent eikonal function.
The Born term is given by
\begin{equation}
\sigma^B_{q{\bar q} \rightarrow Q {\bar Q}}= \pi\alpha_s^2(\mu_R^2) 
K_{q\bar{q}} N_c C_F \left[ \frac{t_1^2 + u_1^2}{s^2} +\frac{2m^2}{s}\right]\,,
\end{equation}
where $K_{q\bar{q}}=N_c^{-2}$ is a color average factor.
In our calculation for the expansion, we used the result for 
the soft matrix at lowest order
in Eq. (A5) of Ref. \cite{Kidonakis:2001gi}. The
lowest-order hard matrix has only one non-zero element, given by
$H_{22}^{q{\bar q}\rightarrow Q {\bar Q}}
=[2/(N_c C_F)] \sigma^B_{q{\bar q} \rightarrow Q {\bar Q}}$.

To be sure, the expansion of the NLL resummed cross section
does not give all the $\delta(s_4)$ terms, only those $\delta(s_4)$ 
terms involving the scale; these terms are 
$\sigma^B_{q{\bar q} \rightarrow Q {\bar Q}} (\alpha_s(\mu_R^2)/\pi)
\delta(s_4) [(-3/2 \newline {}+\ln(t_1u_1/m^4))C_F\ln(\mu_F^2/m^2) 
+(\beta_0/2)\ln(\mu_R^2/m^2)]$.
The rest are obtained by simply matching
with the NLO cross section in \cite{Beenakker:1991ma}. 
Thus we obtain all the $S+V$ terms at NLO.
As shown in Refs. \cite{Meng:1990rp,Kidonakis:1995uu}
these terms dominate the cross section and are an excellent 
approximation at the partonic level to the exact NLO cross section 
close to threshold and even quite far from it.
We note that the exact NLO cross section is the sum of the 
full $S+V$ terms and hard gluon corrections; the latter are not
taken into account by threshold studies and vanish at threshold.

In the DIS scheme, the corresponding result for the $q{\bar q}$ channel is
\begin{eqnarray}
{\hat \sigma}^{\rm DIS \, (1)}_{q{\bar q}\rightarrow Q{\bar Q}}
(s_4,m^2,s,t_1,u_1,\mu_F,\mu_R)&=&\sigma^B_{q{\bar q}\rightarrow Q{\bar Q}}
\frac{\alpha_s(\mu_R^2)}{\pi}
\left\{2C_F\left[\frac{\ln(s_4/m^2)}{s_4}\right]_{+}\right.
\nonumber \\ && \hspace{-45mm} \left.
{}+\left[\frac{1}{s_4}\right]_{+} \left[2 {\rm Re} {\Gamma'}_{22}^{q{\bar q}}
-\frac{C_F}{2}+C_F\ln\left(\frac{s^2}{t_1u_1}\right)
-2C_F\ln\left(\frac{\mu_F^2}{m^2}\right)\right]\right\}
\nonumber \\ && \hspace{-45mm} 
{}+\delta(s_4) \, \sigma^{(1) \, q{\bar q} \, {\rm S+V}}_{\rm DIS} \, ,
\end{eqnarray}
where $\sigma^{(1) \, q{\bar q} \, {\rm S+V}}_{\rm DIS}$
can be obtained from Eq. (4.14) in  \cite{Beenakker:1991ma}.   

For the $gg$ channel in the $\overline{\rm MS}$ scheme the NLO threshold
corrections are given by
\beqa
{\hat \sigma}^{\overline {\rm MS} \, (1)}_{gg \rightarrow Q{\bar Q}}
(s_4,m^2,s,t_1,u_1,\mu_F,\mu_R)&=&
\sigma^B_{gg\rightarrow Q{\bar Q}} \frac{\alpha_s(\mu_R^2)}{\pi} 
\left\{4C_A\left[\frac{\ln(s_4/m^2)}{s_4}\right]_{+}
-2C_A \ln\left(\frac{\mu_F^2}{m^2}\right) 
\left[\frac{1}{s_4}\right]_{+}\right\}
\nonumber \\ && \hspace{-55mm}
{}+\alpha_s^3(\mu_R^2) K_{gg} B_{QED} \left[\frac{1}{s_4}\right]_{+}
\left\{N_c(N_c^2-1)\frac{(t_1^2+u_1^2)}{s^2}
\left[\left(-C_F+\frac{C_A}{2}\right)
{\rm Re} L_{\beta}\right. \right.
\nonumber \\ && \hspace{-20mm} \left.
{}+\frac{C_A}{2}\ln\left(\frac{m^2s}{t_1u_1}\right)
-C_F\right]+\frac{(N_c^2-1)}{N_c}(C_F-C_A) {\rm Re} L_{\beta}
\nonumber \\ && \hspace{-55mm} \left.
{}+C_F \frac{(N_c^2-1)}{N_c}
+\frac{N_c^2}{2}(N_c^2-1)
\ln\left(\frac{u_1}{t_1}\right)\frac{(t_1^2-u_1^2)}{s^2} \right\} 
+\delta(s_4) \, \sigma^{(1) \, gg \, {\rm S+V}}_{{\overline {\rm MS}}} \, ,
\eeqa
where $K_{gg}=(N_c^2-1)^{-2}$ is a color average factor, and
\beq
B_{\rm QED}=\frac{t_1}{u_1}+\frac{u_1}{t_1}+\frac{4m^2s}{t_1u_1}
\left(1-\frac{m^2s}{t_1u_1}\right) \, .
\eeq
Here, $\sigma^{(1)\, gg \, {\rm S+V}}_{{\overline {\rm MS}}}$ 
again denotes the
soft plus virtual $\delta(s_4)$ terms in the NLO cross section. 
These terms are given by Eq. (6.19) in Ref. \cite{Beenakker:1989bq}; 
we note, however, that in that reference the scale was set equal to $m$, 
therefore in addition to those terms 
we have to include in $\sigma^{(1)\, gg \, {\rm S+V}}_{{\overline {\rm MS}}}$ 
the terms $\sigma^B_{gg \rightarrow Q {\bar Q}} (\alpha_s(\mu_R^2)/\pi)
[C_A \ln(t_1u_1/m^4)\, \ln(\mu_F^2/m^2)+(\beta_0/2)\ln(\mu_R^2/\mu_F^2)]$.
The Born term is given by 
\beq
\sigma^B_{gg\rightarrow Q {\bar Q}} =  2\pi \alpha_s^2(\mu_R^2) K_{gg}
N_c C_F \left[C_F - C_A \frac{t_1u_1}{s^2}\right] B_{\rm QED} \, .
\eeq
In our calculation for the expansion we used the result for 
the soft matrix at lowest order
in Eq. (C3) of Ref. \cite{Kidonakis:2001gi}. The
lowest-order hard matrix has the form of Eq. (C6) of 
Ref. \cite{Kidonakis:2001gi}, with independent elements
$H_{11}^{gg\rightarrow Q {\bar Q}}=
\pi \alpha_s^2 B_{\rm QED} K_{gg}/(2N_c^2)$,
$H_{13}^{gg\rightarrow Q {\bar Q}}=N_c H_{11}^{gg\rightarrow Q {\bar Q}}
(t_1^2-u_1^2)/s^2$,
and $H_{33}^{gg\rightarrow Q {\bar Q}}=
N_c^2 H_{11}^{gg\rightarrow Q {\bar Q}} (1-4t_1u_1/s^2)$.

Again as shown in Refs. \cite{Meng:1990rp,Kidonakis:1995uu}
these corrections dominate the cross section near threshold 
and are an excellent approximation at the partonic level to the exact 
NLO cross section. For NLO expansions in PIM kinematics see Appendix C.

\subsection{NNLO-NNLL threshold corrections 
for $q \bar{q} \rightarrow Q \bar{Q}$}

Next we derive the NNLO-NNLL threshold corrections from the 
two-loop expansion of the resummed cross section.

For the $q{\bar q}$ channel in the $\overline {\rm MS}$ scheme these
corrections are 
\beqa
{\hat \sigma}^{\overline {\rm MS} \, (2)}_{q{\bar q}\rightarrow Q{\bar Q}}
(s_4,m^2,s,t_1,u_1,\mu_F,\mu_R)&=&
\sigma^B_{q{\bar q}\rightarrow Q{\bar Q}} 
\left(\frac{\alpha_s(\mu_R^2)}{\pi}\right)^2 
\left\{8 C_F^2 \left[\frac{\ln^3(s_4/m^2)}{s_4}\right]_{+} \right.
\nonumber \\ && \hspace{-55mm}
{}+\left[\frac{\ln^2(s_4/m^2)}{s_4}\right]_{+} \left[-\beta_0 C_F 
+12 C_F\left({\rm Re}{\Gamma'}_{22}^{q{\bar q}}
-C_F+C_F\ln\left(\frac{sm^2}{t_1u_1}\right)
-C_F\ln\left(\frac{\mu_F^2}{m^2}\right)\right)\right]
\nonumber \\ && \hspace{-55mm}
{}+\left[\frac{\ln(s_4/m^2)}{s_4}\right]_{+}
\left[4\left[{\rm Re} {\Gamma'}_{22}^{q{\bar q}}- C_F
-C_F \ln\left(\frac{t_1u_1}{sm^2}\right)
-C_F \ln\left(\frac{\mu_F^2}{m^2}\right)\right]^2 \right.
\nonumber \\ && \hspace{-55mm}
{}+4{\Gamma'}_{12}^{q{\bar q}}{\Gamma'}_{21}^{q{\bar q}}
-\beta_0\left[{\rm Re} {\Gamma'}_{22}^{q{\bar q}}-C_F 
-C_F \ln\left(\frac{t_1 u_1}{sm^2}\right)
-C_F\ln\left(\frac{\mu_R^2}{m^2}\right)\right]
\nonumber \\ && \hspace{-55mm} \left. \left.
{}+2C_F K -16 \zeta_2 C_F^2
+4 C_F \, c^{(1) \, q {\bar q} \, {\rm S+V}}_{{\overline {\rm MS}}}
\right]\right\}
+{\cal O}\left(\left[\frac{1}{s_4}\right]_+\right)\, ,
\label{qqNNLO}
\eeqa
where $c^{(1)\, q {\bar q} \, {\rm S+V}}_{{\overline {\rm MS}}}$ is defined by
\beq
\sigma^{(1) \, q {\bar q} \, {\rm S+V}}_{{\overline {\rm MS}}}
=\frac{\alpha_s}{\pi} \sigma^B_{q {\bar q} \rightarrow Q {\bar Q}} \, 
c^{(1) \, q {\bar q} \, {\rm S+V}}_{{\overline {\rm MS}}} \, ,
\label{c1}
\eeq
and the off-diagonal elements of the soft anomalous dimension matrix 
(dropping an overall $\alpha_s/\pi$) are
\beq
{\Gamma'}_{21}^{q{\bar q}}=2\ln\left(\frac{u_1}{t_1}\right) \, , \quad
{\Gamma'}_{12}^{q{\bar q}}=\frac{C_F}{C_A} \ln\left(\frac{u_1}{t_1}\right)\,.
\eeq
We are able to derive all the NNLL terms by matching with the $S+V$ terms
in the NLO cross section, Eq. (\ref{c1}).

In addition, we can derive at NNLL accuracy the following 
$[1/s_4]_+$ and $\delta(s_4)$ terms involving logarithms of the
factorization and renormalization scales:
\beqa
&& \sigma^B_{q{\bar q}\rightarrow Q{\bar Q}}  
\left(\frac{\alpha_s(\mu_R^2)}{\pi}\right)^2 
\left[\frac{1}{s_4}\right]_+\left\{
\ln^2\left(\frac{\mu_F^2}{m^2}\right)
C_F\left[C_F\left(3-2\ln\left(\frac{t_1u_1}{m^4}\right)\right)
+\frac{\beta_0}{4}\right]\right. 
\nonumber \\ &&
{}-\frac{3}{2} C_F \beta_0 
\ln\left(\frac{\mu_R^2}{m^2}\right)
\ln\left(\frac{\mu_F^2}{m^2}\right)
+\ln\left(\frac{\mu_F^2}{m^2}\right)
\left[-2C_F {\hat T}_{\overline {\rm MS}}^{(1)\, q{\bar q}}
-C_FK+8C_F^2\zeta_2 \right.
\nonumber \\ && \left.
{}+C_F\left(2\ln\left(\frac{t_1u_1}{m^4}\right)-3\right)
\left({\rm Re}{\Gamma'}_{22}^{q{\bar q}}-C_F
+C_F\ln\left(\frac{sm^2}{t_1u_1}\right)\right)\right]
\nonumber \\ && \left.
{}+\frac{3}{2}\beta_0 \ln\left(\frac{\mu_R^2}{m^2}\right)
\left[ {\rm Re}{\Gamma'}_{22}^{q{\bar q}}
-C_F-C_F \ln\left(\frac{t_1u_1}{m^2s}\right)
\right]\right\} 
\nonumber \\ && 
{}+\sigma^B_{q{\bar q}\rightarrow Q{\bar Q}}
\left(\frac{\alpha_s(\mu_R^2)}{\pi}\right)^2
\delta(s_4)\left\{\ln^2\left(\frac{\mu_F^2}{m^2}\right)
\left[\frac{1}{2}C_F^2\ln^2\left(\frac{t_1u_1}{m^4}\right)
+\frac{9}{8}C_F^2-\frac{3}{2}C_F^2\ln\left(\frac{t_1u_1}{m^4}\right)
\right. \right.
\nonumber \\ && \left.
{}-2C_F^2\zeta_2
-\frac{\beta_0}{8}C_F\ln\left(\frac{t_1u_1}{m^4}\right)
+\frac{3}{16}\beta_0 C_F \right]
+\frac{3}{16} \beta_0^2\ln^2\left(\frac{\mu_R^2}{m^2}\right)
\nonumber \\ && \left.
{}+\frac{3}{4}C_F\beta_0
\ln\left(\frac{\mu_F^2}{m^2}\right)\ln\left(\frac{\mu_R^2}{m^2}\right)
\left[\ln\left(\frac{t_1u_1}{m^4}\right)-\frac{3}{2}\right]\right\}\, ,
\label{scaleqqms}
\eeqa
where ${\hat T}_{\overline {\rm MS}}^{(1)\, q{\bar q}}$ is obtained from 
$c^{(1) \,q {\bar q} \, {\rm S+V}}_{{\overline {\rm MS}}}$
in Eq. (\ref{c1}) by dropping all scale terms $\ln(\mu_F/m)$ and
$\ln(\mu_R/m)$ in $c^{(1)\, q {\bar q} \, {\rm S+V}}_{{\overline {\rm MS}}}$.
We note that at NNLL accuracy we derive all $[1/s_4]_+$ scale terms,
but in the $\delta(s_4)$ coefficient we can only determine quadratic 
terms in the scale logarithms. 

\begin{figure}
\centerline{
\psfig{file=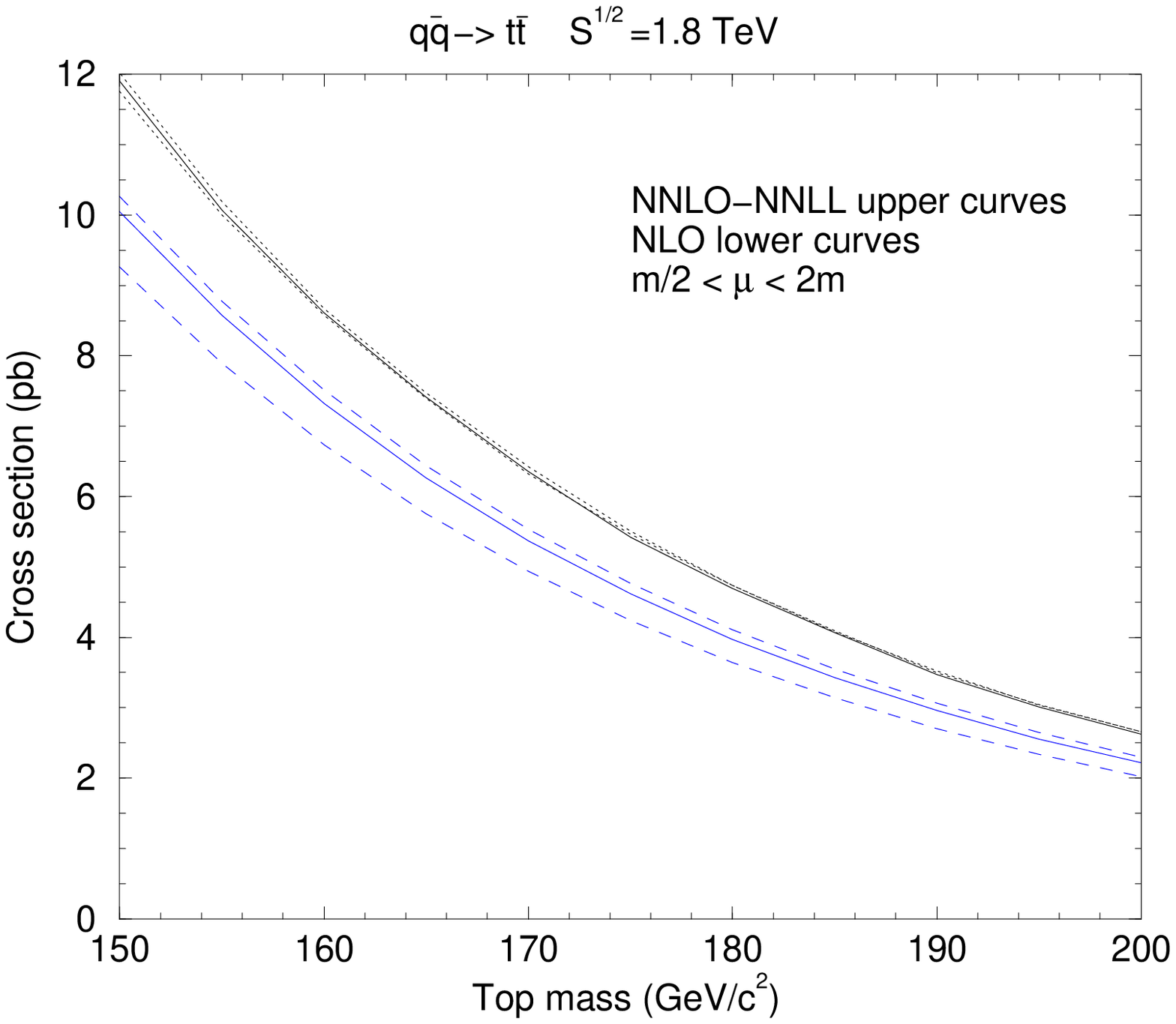,height=3.8in,width=4.5in,clip=}}
{Fig. 1.  Top quark production at the Tevatron  with 
$\sqrt{S} = 1.8$ TeV for the $q{\bar q} \rightarrow t {\bar t}$ 
channel in the ${\overline {\rm MS}}$ scheme. 
Plotted are the exact NLO cross section 
for $\mu=m$ (lower solid line), $m/2$ and $2m$ 
(upper and lower dashed lines), and the NNLL-NNLO cross section
for $\mu=m$ (upper solid line), $m/2$ and $2m$ 
(upper and lower dotted lines).}
\label{fig1}
\end{figure}

Since the $q{\bar q}$ channel is dominant for top quark production at the
Tevatron, it is worthwhile to show some numerical results for the hadronic
cross section in that channel alone before presenting the full cross section
(for details of the hadronic calculation see Appendix B).
In Fig. 1 we plot the NNLO-NNLL top quark cross section 
$q{\bar q} \rightarrow t {\bar t}$ at the Tevatron
with $\sqrt{S}=1.8$ TeV together with the exact NLO cross section
\cite{Beenakker:1991ma,Nason:1988xz} 
as a function of the top mass; we have set $\mu\equiv\mu_F=\mu_R$.
Here and in the rest of the 
paper we use the CTEQ5M parton densities \cite{Lai:2000wy}
when calculating ${\overline{\rm MS}}$ results.
We note the significant increase of the cross section at NNLO
along with the dramatic reduction in scale variation between 
$m/2$ and $2m$. This reduction is also evident for a wide range of scale
choices in Fig. 2. The NNLO cross section is larger than at NLO
and relatively flat with respect to scale variations.

\begin{figure}
\centerline{
\psfig{file=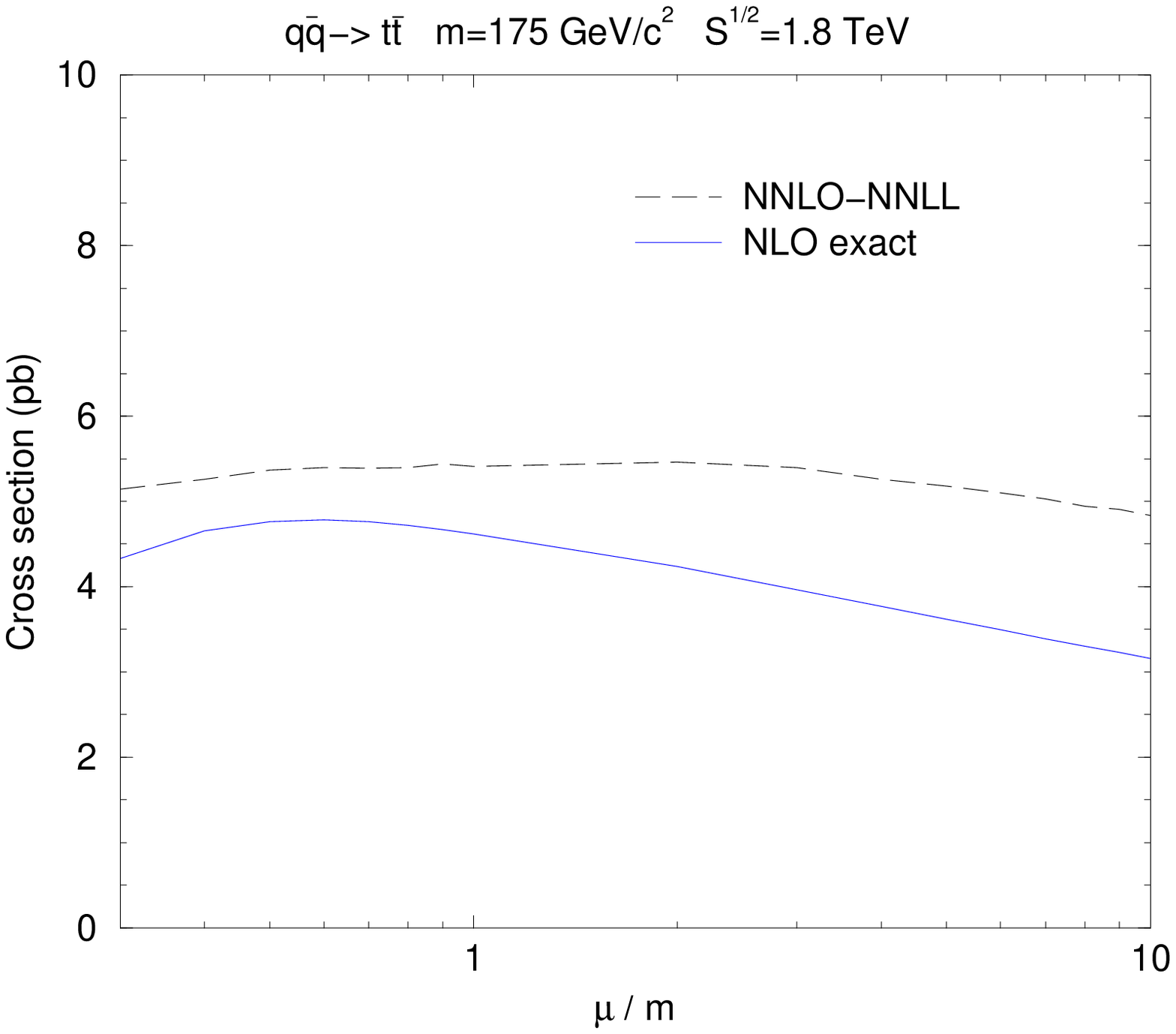,height=3.8in,width=4.5in,clip=}}
{Fig. 2.  The scale dependence of the cross section for 
$q{\bar q} \rightarrow t {\bar t}$ in the
${\overline {\rm MS}}$ scheme at the Tevatron with 
$\sqrt{S} = 1.8$ TeV and $m=175$ GeV/$c^2$.}
\label{fig2}
\end{figure}

In Table 1 we present detailed numerical results for the $q{\bar q}$
corrections to the top quark cross section at the Tevatron 
with $m=175$ GeV/$c^2$ 
through NNLO at both NLL and NNLL accuracy. Some of these numbers will be 
useful in our discussion in the next subsection.

\begin{table}[htb]
\begin{center}
\begin{tabular}{|c|c|c|c|} \hline 
$q{\bar q}\rightarrow t{\bar t}$ & $\mu=m$ & $\mu=m/2$  & $\mu=2m$ \\ \hline
Born  & 3.81  &  5.30 & 2.87  \\ \hline 
NLO-exact & 0.81 & -0.53 & 1.37 \\ \hline
NLO-NLL & 1.31  & 0.03  & 1.81    \\ \hline
NLO-full $S+V$ & 1.26 &  -0.06 & 1.78   \\ \hline
NNLO-NLL & 1.01 & 0.67  & 1.29 \\ \hline
NNLO-NNLL & 0.80 & 0.62 & 1.22 \\ \hline
\end{tabular}
\caption[]{The ${\overline {\rm MS}}$ corrections for top quark
production in the $q {\bar q}$ channel
in pb for $p \overline p$ collisions 
with $\sqrt{S} = 1.8$ TeV and $m=175$ GeV/$c^2$.
Here $\mu=\mu_F=\mu_R$.}
\end{center}
\end{table}

The NNLO-NNLL threshold corrections in the DIS scheme are
\beqa
{\hat \sigma}^{\rm DIS \, (2)}_{q{\bar q}\rightarrow Q{\bar Q}}
(s_4,m^2,s,t_1,u_1,\mu_F,\mu_R)&=&
\sigma^B_{q{\bar q}\rightarrow Q{\bar Q}} 
\left(\frac{\alpha_s(\mu_R^2)}{\pi}\right)^2 
\left\{2 C_F^2 \left[\frac{\ln^3(s_4/m^2)}{s_4}\right]_{+} \right.
\nonumber \\ && \hspace{-55mm}
{}+\left[\frac{\ln^2(s_4/m^2)}{s_4}\right]_{+} \left[-\frac{3\beta_0}{4}C_F 
+6C_F \left({\rm Re} {\Gamma'}_{22}^{q{\bar q}}-\frac{C_F}{4}
+\frac{C_F}{2}\ln\left(\frac{s^2}{t_1u_1}\right)
-C_F\ln\left(\frac{\mu_F^2}{m^2}\right)\right) \right.
\nonumber \\ && \hspace{-55mm}
{}+\left[\frac{\ln(s_4/m^2)}{s_4}\right]_{+}
\left[4\left[{\rm Re} {\Gamma'}_{22}^{q{\bar q}}- \frac{C_F}{4}
-\frac{C_F}{2} \ln\left(\frac{t_1u_1}{s^2}\right)
-C_F \ln\left(\frac{\mu_F^2}{m^2}\right)\right]^2 \right.
\nonumber \\ && \hspace{-30mm} 
{}+4{\Gamma'}_{12}^{q{\bar q}}{\Gamma'}_{21}^{q{\bar q}}
-\beta_0\left[{\rm Re} \Gamma'_{22}-\frac{5}{8}C_F
-\frac{3}{4}C_F \ln\left(\frac{t_1u_1}{m^4}\right)
-\frac{1}{2}C_F \ln\left(\frac{\mu_R^2}{s}\right)\right] 
\nonumber \\ && \hspace{-30mm} \left. \left.
{}+C_F K -4 \zeta_2 C_F^2
+2C_F \, c^{(1) \, q {\bar q} \, {\rm S+V}}_{\rm DIS} 
\right]\right\}+{\cal O}\left(\left[\frac{1}{s_4}\right]_+\right)\, ,
\eeqa
where $c^{(1) \, q {\bar q} \, {\rm S+V}}_{\rm DIS}$ is defined in analogy
to Eq. (\ref{c1}).
As for the ${\overline {\rm MS}}$ corrections, we can also derive additional 
$[1/s_4]_+$ and $\delta(s_4)$ terms involving the scale in the DIS
scheme. These terms are
\beqa
&& \sigma^B_{q{\bar q}\rightarrow Q{\bar Q}}  
\left(\frac{\alpha_s(\mu_R^2)}{\pi}\right)^2 
\left[\frac{1}{s_4}\right]_+\left\{
\ln^2\left(\frac{\mu_F^2}{m^2}\right)
C_F\left[C_F\left(3-2\ln\left(\frac{t_1u_1}{m^4}\right)\right)
+\frac{\beta_0}{4}\right]\right. 
\nonumber \\ &&
{}-\frac{3}{2} C_F \beta_0 
\ln\left(\frac{\mu_R^2}{m^2}\right)
\ln\left(\frac{\mu_F^2}{m^2}\right)
+\ln\left(\frac{\mu_F^2}{m^2}\right)
\left[-2C_F {\hat T}_{\rm DIS}^{(1)\, q{\bar q}}-C_FK+4C_F^2\zeta_2 \right. 
\nonumber \\ && \left.
{}+C_F\left(\ln\left(\frac{t_1u_1}{m^4}\right)-\frac{3}{2}\right)
\left(2{\rm Re}{\Gamma'}_{22}^{q{\bar q}}-\frac{C_F}{2}
+C_F\ln\left(\frac{s^2}{t_1u_1}\right)\right)\right]
\nonumber \\ && \left.
{}+\frac{3}{2}\beta_0 \ln\left(\frac{\mu_R^2}{m^2}\right)
\left[{\rm Re}{\Gamma'}_{22}^{q{\bar q}}
-\frac{C_F}{4}-\frac{C_F}{2} \ln\left(\frac{t_1u_1}{s^2}\right)
\right]\right\}
\nonumber \\ && 
{}+\sigma^B_{q{\bar q}\rightarrow Q{\bar Q}}
\left(\frac{\alpha_s(\mu_R^2)}{\pi}\right)^2
\delta(s_4)\left\{\ln^2\left(\frac{\mu_F^2}{m^2}\right)
\left[\frac{1}{2}C_F^2\ln^2\left(\frac{t_1u_1}{m^4}\right)
+\frac{9}{8}C_F^2-\frac{3}{2}C_F^2\ln\left(\frac{t_1u_1}{m^4}\right)
\right. \right.
\nonumber \\ && \left.
{}-2C_F^2\zeta_2-\frac{\beta_0}{8}C_F\ln\left(\frac{t_1u_1}{m^4}\right)
+\frac{3}{16}\beta_0 C_F \right]
+\frac{3}{16} \beta_0^2\ln^2\left(\frac{\mu_R^2}{m^2}\right)
\nonumber \\ && \left.
{}+\frac{3}{4}C_F\beta_0
\ln\left(\frac{\mu_F^2}{m^2}\right)\ln\left(\frac{\mu_R^2}{m^2}\right)
\left[\ln\left(\frac{t_1u_1}{m^4}\right)-\frac{3}{2}\right]\right\}\, ,
\label{scaleqqdis}
\eeqa
where ${\hat T}_{\rm DIS}^{(1)\, q{\bar q}}$ is defined in analogy to
its $\overline {\rm MS}$ counterpart.

At the Tevatron, with $\sqrt{S}=1.8$ TeV and 
$\mu_F=\mu_R=m=175$ GeV/c$^2$, and with the 
CTEQ5D parton densities \cite{Lai:2000wy},
the exact NLO cross section in the DIS scheme is 4.60 pb and the 
NNLO-NNLL corrections  provide an additional 0.30 pb. The NNLO corrections 
in the DIS scheme
are much smaller than the corresponding ${\overline {\rm MS}}$ corrections;
that follows from the definition of the two schemes.

Results for the NNLO expansion of the resummed cross section in PIM kinematics 
in both the ${\overline {\rm MS}}$ and DIS schemes are presented
in Appendix C (see also  Ref. \cite{Kidonakis:2000ze}).
The differences between the expansions in the two different kinematics 
are in the extra terms involving $\ln(t_1u_1/m^4)$ for the 1PI fixed-order 
expansions relative to the PIM expansions and in the 
matching terms needed to reach NNLL accuracy \cite{KLMV}.

\subsection{Subleading logarithms and resummation prescriptions}

In the previous subsection we derived all the NNLO threshold
corrections for $q {\bar q} \rightarrow t {\bar t}$
through NNLL accuracy. It is interesting to study the effect
of subleading $[1/s_4]_+$ and $\delta(s_4)$ terms that come about when 
inverting the cross section from moment to momentum space.
As we will see, this is intimately related to the disagreements between
various resummation prescriptions that have been proposed.
These prescriptions are needed to avoid the soft gluon divergences 
in the resummed cross section that appear when $\alpha_s$
reaches the Landau pole. As we have discussed before,
in a finite-order expansion, as presented in this paper,
there are no divergences and the results are prescription independent.

There are three resummation prescriptions available in the literature.
The earliest is the $x$-space formalism of  
Refs. \cite{Laenen:1992af,Laenen:1994xr}.
The resummation was performed at leading logarithmic (LL) accuracy 
in momentum space and a cutoff was chosen
to avoid the divergence. In practice, the cutoff was chosen so that 
numerically the resummed result would agree with the expansion of the 
resummed cross section through NNLO. 
This approach was also used at NLL accuracy in 
Refs. \cite{Kidonakis:1997zd,Kidonakis:1997ed,Kidonakis:1998ei}.
The finite-order expansion is essentially the same whether the resummation
is performed in momentum or moment space so, although the approach
is quite different, in practice the numerical results from this
approach are not inconsistent with the ones we are presenting here at LL and
NLL accuracy.

The second prescription is principal value resummation, originally
developed for Drell-Yan production in Ref. \cite{Contopanagos:1994yq}. 
A principal value prescription is used to bypass the Landau pole. 
This approach was used at LL accuracy for top quark production 
in Refs. \cite{Berger:1996ad,Berger:1998gz}. Numerically
the results are similar to those of Ref. \cite{Laenen:1992af}. 
This approach has not yet been used at NLL accuracy for top quark production 
at present.

The third approach is the minimal prescription of Ref. \cite{Catani:1996yz}.
It has been applied at both LL \cite{Catani:1996yz} and 
NLL accuracy \cite{Bonciani:1998vc} to heavy quark
production. Numerically it differs substantially from the results
of Refs. \cite{Laenen:1992af,Berger:1996ad}. 
As discussed in Refs. \cite{Catani:1996yz,Berger:1998gz} 
this difference emerges from 
extra subleading terms, which are kept in the minimal prescription
approach, that come  from the inversion of the 
resummed cross section from moment to momentum space.

Let us begin our study of subleading terms by rewriting the 
${\overline {\rm MS}}$ NLO corrections for $q {\bar q} \rightarrow Q {\bar Q}$
in Eq. (\ref{NLOqq}) in the shorthand notation
\beq
{\hat\sigma}^{(1)}(s_4)=\sigma^B \frac{\alpha_s}{\pi} \left\{c_1 \delta(s_4)
 +c_2 \left[\frac{1}{s_4}\right]_+
+c_3 \left[\frac{\ln(s_4/m^2)}{s_4}\right]_+\right\}\, ,
\label{NLOb}
\eeq
with $c_3=4C_F$, $c_2=2 {\rm Re} {\Gamma'}_{22}^{q{\bar q}}
-2C_F+2C_F\ln(sm^2/(t_1u_1))-2C_F\ln(\mu_F^2/m^2)$, and
$c_1=c^{(1) \, q {\bar q} \, {\rm S+V}}_{{\overline {\rm MS}}}$.
This result actually comes from the inversion of the moment space expression
\beq
{\hat\sigma}^{(1)}(N)=\sigma^B \frac{\alpha_s}{\pi} \left\{c_1
+c_2 I_0(N) +c_3 I_1(N)\right\}
=\sigma^B \frac{\alpha_s}{\pi} \left\{c_1
 -c_2 \ln {\tilde N} +\frac{c_3}{2} (\ln^2 {\tilde N}+\zeta_2)\right\} \, ,
\label{NLOmom}
\eeq
where ${\tilde N}=N e^{\gamma_E}$, with $\gamma_E$ the Euler constant,
and $I_0$ and $I_1$ are given in Appendix A.
Eq. (\ref{NLOmom}) comes directly from the 
expansion of the resummed cross section in moment space in Eq. (\ref{resHQ}).

Now, let us examine the NLO expansion at NLL accuracy, with $\mu_F=\mu_R=m$. 
At that accuracy, we keep only the $\ln^2 N$ and $\ln N$ terms
in Eq. (\ref{NLOmom}). Since $\ln N=\ln {\tilde N} -\gamma_E$,
upon inversion to momentum space we get back the
$[\ln(s_4/m^2)/s_4]_+$ and $[1/s_4]_+$ terms in Eq. (\ref{NLOb})
plus the following extra terms: 
\beq
\sigma^B \frac{\alpha_s}{\pi}\left[-\frac{c_3 \gamma_E^2}{2}
+c_2 \gamma_E -\frac{\zeta_2}{2} c_3\right] \, \delta(s_4) \, .
\label{1unphys}
\eeq
But there are no terms involving $\gamma_E$ in the exact NLO calculation,
i.e. in the term $c_1$ in Eq. (\ref{NLOb}) which comes from  
$\sigma^{(1) \, q {\bar q} \, {\rm S+V}}_{{\overline {\rm MS}}}$;
therefore these terms are clearly unphysical. They should not appear 
in the cross section because
of the definition of the ${\overline {\rm MS}}$ scheme.
The $\gamma_E$ terms are an artifact of the inversion from moment to 
momentum space. Indeed, if we had kept NLL terms in $\ln {\tilde N}$
rather than $\ln N$, there would be no $\gamma_E$ terms. 
Also the coefficient of the $\zeta_2$ term is wrong. 
As can be seen from the full NLO corrections, it has the wrong sign. 
We can study the numerical effect of these extra terms on the cross section.
At the Tevatron, with $\sqrt{S}=1.8$ TeV and $m=175$ GeV/c$^2$,
the full NLO $S+V$ corrections,
Eq. (\ref{NLOqq}), are 1.26 pb for $\mu_F=\mu_R=m$, see Table 1. 
At NLL accuracy, the corrections are 1.31 pb.
If we include all the unphysical terms of Eq. (\ref{1unphys}),  the NLL
corrections become 0.39 pb, clearly very far from the true size of 
the full $S+V$ corrections.
Keeping only the $\zeta_2$ term in Eq. (\ref{1unphys}) we find 0.74 pb,
which is closer to  but still well below the full $S+V$ corrections.
If we keep only the $\zeta_2$ term but with the opposite sign, as in the full
corrections, we find 1.88 pb. It is clear that keeping 
unphysical subleading terms in the cross section 
can produce very misleading results.
Even if we discard the unphysical terms and keep only some of the 
physical subleading terms 
we can still make erroneous predictions, especially if the 
coefficients are wrong.
One of the greatest achievements (or, from another viewpoint, tests)
of the formalism of theshold resummation at NLL and higher accuracy
is that it accurately reproduces the exact NLO cross section both
analytically and numerically. In fact, one may argue that 
only because of this agreement is threshold resummation worthwhile.
After all, if the corrections to be resummed are not dominant
then the necessity for resummation is greatly diminished.
That holds not only for heavy quark production but also for 
other QCD processes such as direct photon \cite{Kidonakis:2000hq}, 
W+jet \cite{Kidonakis:2000ur}, and single-jet production 
\cite{Kidonakis:2001gi}. Therefore it is important to ensure that
we don't introduce terms, unphysical or otherwise, that would spoil 
this agreement.

We can extend our study of subleading logarithms to NNLO.
The NNLO threshold corrections in moment space are given in shorthand 
notation by
\beqa
{\hat\sigma}^{(2)}(N)&=&\sigma^B\frac{\alpha_s^2}{2\pi^2}
\left\{\frac{c_3^2}{4} \ln^4{\tilde N}-c_3 c_2 \ln^3{\tilde N}
+\left[c_3\left(c_1+\frac{1}{2}\zeta_2 c_3\right)
+c_2^2\right] \ln^2{\tilde N} \right.
\nonumber \\ && \left.
{}-2 c_2 \ln{\tilde N} \left(c_1+\frac{1}{2}\zeta_2 c_3\right)
+\left(c_1+\frac{1}{2}\zeta_2 c_3\right)^2 
+{\tilde F}(\beta_0, \Gamma_S^2, K, 2-{\rm loop}) \right\} \, ,
\label{NNLOms}
\eeqa
i.e. by the square of the terms in curly brackets in Eq. (\ref{NLOmom}),
plus a function $\tilde F$ that comprises the $\beta_0$ terms 
that come from changing the argument in the running coupling,
$\alpha_s(\mu'^2)=\alpha_s(\mu^2)[1-\beta_0 \ln(\mu'^2/\mu^2)
(\alpha_s(\mu^2)/(4\pi)]$;
the two-loop $K$ terms, with $K$ defined below Eq. (\ref{Eexp});
square terms from the off-diagonal soft anomalous dimension matrix elements;
and two-loop $\Gamma_S$ and other terms (we have also absorbed in $\tilde F$
a term $-T_1^2$, with $T_1=c_1(\mu=m)$). Note that apart from the
$K$ terms, the other two-loop terms appear only beyond NNLL
accuracy and are not known at present.
We then rewrite Eq. (\ref{NNLOms}) in terms of $I_3,I_2,I_1$, 
and $I_0$ defined in Appendix A. 

We can then immediately invert back to momentum space and find
\beqa
{\hat\sigma}^{(2)}(s_4)&=&\sigma^B\frac{\alpha_s^2}{\pi^2}
\left\{\frac{1}{2}c_3^2 \left[\frac{\ln^3(s_4/m^2)}{s_4}\right]_+
+\frac{3}{2}c_3 c_2 \left[\frac{\ln^2(s_4/m^2)}{s_4}\right]_+ \right.
\nonumber \\ &&
{}+\left(c_3c_1+c_2^2-\zeta_2 c_3^2\right)
\left[\frac{\ln(s_4/m^2)}{s_4}\right]_+
+\left(c_2 c_1-\zeta_2 c_2 c_3+\zeta_3 c_3^2\right) 
\left[\frac{1}{s_4}\right]_+
\nonumber \\ && \left.
{}+\left(\frac{c_1^2}{2}-\frac{c_2^2}{2}\zeta_2
+\frac{1}{4}c_3^2\zeta_2^2+\zeta_3c_3c_2
-\frac{3}{4}\zeta_4 c_3^2\right) \delta(s_4) 
+F(\beta_0, \Gamma_S^2, K, 2-{\rm loop}) \right\} \, .
\nonumber \\
\label{full2l}
\eeqa
Here $F$ denotes the terms that come from the inversion of ${\tilde F}$
and starts contributing at NLL and higher accuracy.
One can easily see that this form, with the appropriate explicit expression 
for $F$, agrees with the NNLO-NNLL expansion
given in Eq. (\ref{qqNNLO}), including the additional scale terms 
of Eq. (\ref{scaleqqms}).
Clearly we don't know all the $[1/s_4]_+$ and $\delta(s_4)$ terms 
because of unknown two-loop corrections in $F$. 
Of course there are no $\gamma_E$ terms in the NNLO cross section.
This is also known from the two-loop Drell-Yan cross section
\cite{Hamberg:1991np,vanNeerven:1992gh,Magnea:1991qg}: many of the
NNLO terms are the same as for top quark production since the exponent
in the resummed cross section that comes from the incoming partons
is universal; and there are no $\gamma_E$ terms. Again, this follows from 
the definition of the ${\overline {\rm MS}}$ scheme.
These terms are clearly unphysical at both NLO and NNLO and 
indeed at any higher order.

Now, let us see what happens if one keeps the logarithms only at a certain
accuracy.
At NNLL accuracy, we keep the $\ln^4 N$, $\ln^3 N$, and $\ln^2 N$ terms 
in ${\hat \sigma}^{(2)}(N)$, Eq. (\ref{NNLOms}). 
Then, the subleading terms from the inversion to momentum space are:
\beqa
&& \sigma^B \frac{\alpha_s^2}{\pi^2} \left\{
\left(\zeta_3 c_3^2 -\frac{3}{2} c_3 c_2 \zeta_2 \right)
\left[\frac{1}{s_4}\right]_{+} \right.
\nonumber \\ && \left.
{}+\delta(s_4) \left(\frac{c_3^2 \zeta_2^2}{8}+c_3 c_2 \zeta_3 
-\frac{3}{4}c_3^2 \zeta_4 -\frac{\zeta_2}{2}c_2^2
-\frac{\zeta_2}{2} c_3 c_1 
-2\zeta_2 {\Gamma'}_{12}^{q{\bar q}}{\Gamma'}_{21}^{q{\bar q}}\right) \right\}
\label{zeta}
\eeqa
plus
\beqa
&& \sigma^B \frac{\alpha_s^2}{\pi^2} \left\{\beta_0 C_F \zeta_2
\left[\frac{1}{s_4}\right]_{+} \right.
\nonumber \\ && \left. 
{}+\delta(s_4)\left[-\frac{2}{3}
\beta_0 C_F \zeta_3-\zeta_2\left(-\frac{\beta_0}{4}c_2
+C_F K\right)\right]\right\}
\label{beta}
\eeqa
plus
\beqa
&& \sigma^B \frac{\alpha_s^2}{\pi^2}  \left\{\gamma_E\left(\beta_0
C_F \gamma_E+4{\Gamma'}_{12}^{q{\bar q}}{\Gamma'}_{21}^{q{\bar q}}
-\frac{\beta_0}{2}c_2
+2C_F K \right)\left[\frac{1}{s_4}\right]_{+} \right.
\nonumber \\ && \left.
{}+\delta(s_4)
\gamma_E^2\left(\frac{2}{3}\beta_0 C_F \gamma_E
+2{\Gamma'}_{12}^{q{\bar q}}{\Gamma'}_{21}^{q{\bar q}}
-\frac{\beta_0}{4}c_2+C_F K\right)\right\}
\nonumber \\ &&
{}+ \sigma^B \frac{\alpha_s^2}{\pi^2} \left\{\left[\frac{c_3^2}{2}\gamma_E^3
-\frac{3}{2}c_3c_2 \gamma_E^2+2 \gamma_E \left(c_3c_1+c_3^2 
\frac{\zeta_2}{2}+c_2^2\right)\right]\left[\frac{1}{s_4}\right]_{+} \right.
\nonumber \\ && \left.
{}+\delta(s_4)\left[\frac{3}{8}c_3^2 \gamma_E^4-c_3c_2\gamma_E^3+\gamma_E^2
\left(c_3c_1+c_3^2\frac{\zeta_2}{2}+c_2^2\right)\right]\right\} \, .
\label{gammaE}
\eeqa
The subleading $\zeta$ terms in Eq. (\ref{zeta}) 
appear also in the full cross section
of Eq. (\ref{full2l}). However, a comparison between these two equations
shows that some of these terms have the wrong coefficients. 
The terms in Eq. (\ref{gammaE}) are again the unphysical $\gamma_E$
terms that should not appear at any order of the perturbative series.  Again,
if we had kept NNLL terms in $\ln {\tilde N}$
rather than $\ln N$, there would be no $\gamma_E$ terms.
The $\beta_0$ and $K$ terms in Eq. (\ref{beta}) 
would also be absent in an exact calculation. This is because
they appear in integrals of the form of Eq. (A.1), which upon inversion 
to momentum space should give back the original ``plus'' distributions with 
no subleading terms.  
 
We can again study the numerical effect of these extra subleading terms.
In the following, we keep $\mu_F=\mu_R=m$. At the Tevatron, with
$\sqrt{S}=1.8$ TeV and $m=175$ GeV/c$^2$, the ${\overline {\rm MS}}$
NNLO-NNLL corrections for $q {\bar q} \rightarrow t {\bar t}$
are 0.80 pb as we saw in Table 1.
If we keep also the subleading $\zeta$ terms in Eq. (\ref{zeta})
the result becomes 0.39 pb. If we keep the subleading terms in both 
Eqs. (\ref{zeta}) and (\ref{beta}), the corrections become 0.13 pb.
Finally, if we include all subleading terms, Eqs. (\ref{zeta}), (\ref{beta}),
and (\ref{gammaE}) the corrections become 0.08 pb.
This last result is similar to the result presented in 
Ref. \cite{Bonciani:1998vc} (note that different parton densities are used;
also our formalism resums the fully differential cross section 
while \cite{Bonciani:1998vc} resums only the total cross section; the latter
approach introduces some additional errors, see the discussion in 
Ref. \cite{Sterman:2001pt}). 
Clearly the  inclusion of the unphysical $\gamma_E$ and other terms 
decreases the NNLO-NNLL corrections by a
factor of ten. The effects of these unphysical terms are much bigger than those
of the LL, NLL, and NNLL terms. It is difficult to accept a result
in which unphysical subleading terms dominate the three leading powers
of the logarithms. Such a result defies the meaning of 
leading level, next-to-leading level and so on. And as was evidenced
by the NLO exercise, a result with these subleading terms substantially
underestimates the correct value for the cross section.
We also note that if we keep the subleading $\zeta$ terms from the inversion
at full accuracy, as in Eq. (\ref{full2l}), the corrections are 0.45 pb,
much closer to the NNLO-NNLL result.
Of course, we can't derive the full NNLO cross section beyond NNLL
accuracy because of missing two-loop terms, but this certainly indicates
that the corrections with subleading terms tend to get larger
the better the accuracy.

We can also repeat this exercise at NLL accuracy. Here we will
disregard the $\gamma_E$ and $\beta_0, K$ terms.
At NLL accuracy for ${\hat \sigma}^{(2)}$, the subleading $\zeta$ terms  
from the inversion are
\beqa
&& \sigma^B \frac{\alpha_s^2}{\pi^2} \left\{-\frac{3}{2}\zeta_2 c_3^2
\left[\frac{\ln(s_4/m^2)}{s_4}\right]_{+}
+\left(\zeta_3 c_3^2 -\frac{3}{2} c_3 c_2 \zeta_2 \right)
\left[\frac{1}{s_4}\right]_{+} \right.
\nonumber \\ && \left.
{}+\delta(s_4) \left[\frac{3}{8}c_3^2 \left(\zeta_2^2-2\zeta_4\right)
+c_3 c_2 \zeta_3 \right]\right\} \, .
\eeqa
Again, the subleading terms above appear also in the full cross section
of Eq. (\ref{full2l}) but some of these terms have the wrong coefficients. 
As we saw in Table 1, the NNLO-NLL corrections are 1.01 pb.
If one adds the subleading terms above, the result becomes 0.23 pb.
Our conclusions remain the same.
We also note that if we keep only the $[(\ln(s_4/m^2))/s_4]_{+}$ 
subleading terms from the inversion, 
we find that the corrections are 0.58 pb, quite
different from the result we get (0.80 pb) when we calculate the full 
NNLL terms (if we also include the unphysical terms the disagreement is 
far worse). In addition, we observe that if we perform a LL calculation 
and then include subleading NLL (i.e. $[(\ln^2(s_4/m^2))/s_4]_{+}$) terms 
from the inversion involving $\gamma_E$,
the corrections become 0.28 pb versus the 1.01 pb that we find with a full
NLL calculation. All these exercises highlight
the numerical problems that are encountered if one includes 
unjustified subleading terms in the expansion.

As is noted in Ref. \cite{Kidonakis:1996hb}, the integrals of
the leading logarithmic distributions with any smooth function,
such as the convolution with parton distributions,
produce factorial contributions at $n$th order
of the form $\alpha_s^n \, (2n-1)!/n!+...$.
These factorial terms naturally arise in both the exact cross section
at any fixed order and in the finite-order expansions of the resummed 
cross section. 
In the minimal prescription of Ref. \cite{Catani:1996yz}
subleading terms are kept in the resummed cross section and its expansion
in order to avoid certain power corrections, arising from these 
factorial contributions,
which have been shown to be absent in the Drell Yan cross section in 
Refs. \cite{Beneke:1995pq,Sterman:1999gz}.
This is indeed a problem that deserves attention as pointed out
in Ref. \cite{Catani:1996yz}. However, the absence of unphysical
power corrections does not require the introduction in the expansion
of unphysical terms, which, as we have seen above, may greatly underestimate 
the true value of the cross section. 
The exact cross section does not have these power corrections
but also it does not have the $\gamma_E$ terms, only the $\zeta$ terms,
as we have seen explicitly at NLO and NNLO.
Moreover, one has to be careful not to introduce extra terms, even 
physical ones, that produce erroneous numerical results. At low
orders, with $n=1,2$ the factorial contributions are negligible or small
anyway, and certainly smaller than other terms in the expansion
(see also discussion in Ref. \cite{Berger:1998gz}). 
At higher orders of course the factorial contribution grows
and moreover we have ever increasing numbers of uknown coefficients
of subleading logarithms (which actually may be more important numerically
than the factorial terms). Therefore we stop the expansion at NNLO,
thus avoiding the theoretical problem with power corrections.
At NLO and NNLO we can trust the perturbative expansion, as we have determined
all logarithmic coefficients. Our numerical analyses above confirm that.
We also note that in our approach the results for the NNLO corrections
do not change substantially when going from LL to NNLL accuracy (they
change from 0.59 pb to 0.80 pb) while the corresponding results with 
all subleading terms change by almost an order of magnitude (that can also
be seen by comparing Table 1 of Ref. \cite{Catani:1996yz} with Table 2 of 
Ref. \cite{Bonciani:1998vc}).
The relative stability of our results versus logarithmic accuracy is an
additional justification of our approach. 
We can investigate keeping the $\zeta$ terms as a rough estimate of error, 
as we have done above, but we should keep in mind that they may not 
necessarily offer an improvement on the calculation as evidenced 
from the NLO results.
Therefore, we find it best in the numerical analyses presented in this
paper not to retain any terms beyond NNLL accuracy.   
Thus, we do not find the very fast convergence of the higher order corrections
that is claimed in Refs. \cite{Catani:1996yz,Bonciani:1998vc}.

In a recent paper \cite{Lai:1999ik} it is argued that threshold enhancements
are dominated by the region where the hierarchy among different  
powers of the threshold logarithms is lost, 
and therefore NLL resummation is not reliable.
Our numerical results at fixed order do not agree with this claim,
although it is certainly true that the coefficients of lower powers of the
logarithm can be large. 
Moreover, we note that even if the hierarchy among different 
powers of the logarithms were lost, 
at NNLO we have determined the coefficients of all the powers 
of the logarithms, so our results are reliable regardless.
Beyond NNLO, however, there are subleading powers of logarithms 
with undetermined coefficients, which can be large, and then the 
ambiguities with regard to the effect of subleading terms increase. 
Therefore, as discussed above, for detailed numerical results we prefer 
to stop the expansion at NNLO.

\subsection{NNLO-NNLL threshold corrections for $gg \rightarrow Q \bar{Q}$}

For the $gg$ channel in the $\overline{\rm MS}$ scheme the 
NNLO-NNLL corrections are
\beqa
{\hat \sigma}^{\overline {\rm MS} \, (2)}_{gg \rightarrow Q{\bar Q}}
(s_4,m^2,s,t_1,u_1,\mu_F,\mu_R)&=&
\sigma^B_{gg\rightarrow Q{\bar Q}} \left(\frac{\alpha_s(\mu_R^2)}{\pi}\right)^2
\left\{8C_A^2\left[\frac{\ln^3(s_4/m^2)}{s_4}\right]_{+}\right.
\nonumber \\ && \hspace{-55mm}\quad \quad \left.
+\left[-\beta_0 C_A -12 C_A^2\ln\left(\frac{\mu_F^2}{m^2}\right)\right]
\left[\frac{\ln^2(s_4/m^2)}{s_4}\right]_{+} \right\}
\nonumber \\ && \hspace{-55mm}
{}+\frac{\alpha_s^4(\mu_R^2)}{\pi} K_{gg} B_{\rm QED} 
\left[\frac{\ln^2(s_4/m^2)}{m^2}\right]_{+} C_A \, 3(N_c^2-1)
\left\{\frac{(t_1^2+u_1^2)}{s^2} \right.
\nonumber \\ && \hspace{-55mm} \quad \times \, 
\left[N_c^2 \ln\left(\frac{m^2s}{t_1u_1}\right)
-2N_c\left(C_F-\frac{C_A}{2}\right){\rm Re}L_{\beta}-2 N_c C_F\right]
+2\frac{C_F}{N_c}
\nonumber \\ && \hspace{-55mm} \quad \left.
{}+2\frac{1}{N_c}(C_F-C_A) \,{\rm Re} L_{\beta}
+N_c^2\frac{(t_1^2-u_1^2)}{s^2}\ln\left(\frac{u_1}{t_1}\right)\right\} 
\nonumber \\ && \hspace{-55mm} 
{}+\left[\frac{\ln(s_4/m^2)}{m^2}\right]_{+}
\left\{\left(\frac{\alpha_s(\mu_R^2)}{\pi}\right)^2
\sigma^B_{gg \rightarrow Q{\bar Q}}
\left[\beta_0\left(C_A+C_A\ln\left(\frac{\mu_R^2}{m^2}\right)
+C_A\ln\left(\frac{t_1 u_1}{m^2s}\right)\right) \right. \right.
\nonumber \\ &&  \hspace{-30mm} \left.
{}+2C_A K-16 \zeta_2 C_A^2
+4 C_A \, c^{(1) \, gg \, {\rm S+V}}_{{\overline {\rm MS}}} \right]
\nonumber \\ && \hspace{-55mm}
{}+\frac{\alpha_s^4(\mu_R^2)}{2 \pi (N_c^2-1)}\, B_{\rm QED}
\left(1-\frac{2 t_1 u_1}{s^2}\right)
\left[4 N_c \left({\rm{Re}} {\Gamma'}_{22}^{gg} - C_A
-C_A\ln\left(\frac{\mu_F^2}{s}\right)-C_A\ln\left(\frac{t_1 u_1}{m^4}\right)
\right)^2 \right.
\nonumber \\ && \hspace{-30mm} \left.
{}+\frac{N_c}{4}(N_c^2+4)({\Gamma'}^{gg}_{31})^2
-\beta_0 N_c {\rm{Re}}{\Gamma'}_{22}^{gg} \right] 
\nonumber \\ && \hspace{-55mm}
{}+\frac{\alpha_s^4(\mu_R^2)}{2 \pi (N_c^2-1)}\, B_{\rm QED}
\left(\frac{t_1^2-u_1^2}{s^2}\right)
\left[4 {\Gamma'}_{31}^{gg} \left({\rm{Re}} {\Gamma'}_{11}^{gg}-C_A
-C_A\ln\left(\frac{\mu_F^2}{s}\right)
-C_A\ln\left(\frac{t_1 u_1}{m^4}\right)\right) \right.
\nonumber \\ && \hspace{-50mm} \left. 
{}+2(N_c^2-2) {\Gamma'}_{31}^{gg} \left({\rm{Re}} {\Gamma'}_{22}^{gg}-C_A
-C_A\ln\left(\frac{\mu_F^2}{s}\right)
-C_A\ln\left(\frac{t_1 u_1}{m^4}\right)\right)
-\beta_0 \frac{N_c^2}{4} {\Gamma'}_{31}^{gg}\right]
\nonumber \\ && \hspace{-55mm} 
{}+\frac{\alpha_s^4(\mu_R^2)}{2 \pi (N_c^2-1)}\, B_{\rm QED}
\left[\frac{4}{N_c} \left({\rm{Re}} {\Gamma'}_{11}^{gg}-C_A
-C_A\ln\left(\frac{\mu_F^2}{s}\right)
-C_A\ln\left(\frac{t_1 u_1}{m^4}\right)\right)^2
-\frac{N_c}{2}({\Gamma'}^{gg}_{31})^2 \right.
\nonumber \\ && \hspace{-50mm} \left. \left.
{}-\frac{8}{N_c} \left({\rm Re} {\Gamma'}_{22}^{gg} -C_A
-C_A\ln\left(\frac{\mu_F^2}{s}\right)
-C_A\ln\left(\frac{t_1 u_1}{m^4}\right)\right)^2
-\beta_0 \frac{1}{N_c} {\rm Re}\left({\Gamma'}_{11}^{gg}
-2 {\Gamma'}_{22}^{gg}\right) \right] \right\}
\nonumber \\ && \hspace{-55mm}
+{}{\cal O}\left(\left[\frac{1}{s_4}\right]_+\right)\, ,
\nonumber \\ 
\eeqa
where $c^{(1) \, gg \, {\rm S+V}}_{\overline {\rm MS}}$ is defined 
in analogy to Eq. (\ref{c1}), and
the elements of the soft anomalous dimension matrix 
(dropping gauge-dependent terms and an overall $\alpha_s/\pi$) are
\begin{eqnarray}
{\rm Re}{\Gamma'}_{11}^{gg}&=&-C_F({\rm Re} L_{\beta}+1)+C_A \,, \quad \quad 
{\Gamma'}_{31}^{gg}=\ln\left(\frac{u_1^2}{t_1^2}\right) \, ,
\nonumber \\
{\rm Re}{\Gamma'}_{22}^{gg}&=&-C_F({\rm Re} L_{\beta}+1)
+\frac{C_A}{2}\left[2+\ln\left(\frac{t_1 u_1}{m^2 s}\right)
+{\rm Re} L_{\beta}\right] \, .
\label{GammaggQQ}
\end{eqnarray}
As for the $q{\bar q}$ channel, we can also derive at NNLL accuracy additional 
$[1/s_4]_+$ and $\delta(s_4)$ terms involving the scale. These terms are
\beqa
&& \hspace{-7mm} \sigma^B_{gg\rightarrow Q{\bar Q}} 
\left(\frac{\alpha_s(\mu_R^2)}{\pi}\right)^2
\left[\frac{1}{s_4}\right]_{+}
\left\{\ln^2\left(\frac{\mu_F^2}{m^2}\right)
C_A \left[\frac{5}{4}\beta_0-2C_A\ln\left(\frac{t_1u_1}{m^4}\right)
\right]-\frac{3}{2} C_A \beta_0 
\ln\left(\frac{\mu_R^2}{m^2}\right) \ln\left(\frac{\mu_F^2}{m^2}\right)
\right.
\nonumber \\ && \quad 
{}+\ln\left(\frac{\mu_F^2}{m^2}\right)
\left[-2C_A {\hat T}^{(1)\, gg}_{\overline {\rm MS}} 
+8 C_A^2 \zeta_2 -C_A K +\beta_0 N_c 
+\beta_0 N_c \ln\left(\frac{t_1 u_1}{m^2 s}\right) \right.
\nonumber \\ && \quad \quad \quad \left.
{}-2 N_c^2 \ln^2\left(\frac{t_1 u_1}{m^4}\right)
-2 N_c^2 \ln\left(\frac{t_1 u_1}{m^4}\right) 
\ln\left(\frac{m^2}{s}\right)
-2 N_c^2 \ln\left(\frac{t_1 u_1}{m^4}\right)\right]
\nonumber \\ && \quad \left.
{}-\frac{3}{2}C_A \beta_0 \ln\left(\frac{\mu_R^2}{m^2}\right)
\left[\ln\left(\frac{t_1 u_1}{m^2 s}\right) +1\right] \right\}
\nonumber \\ && \hspace{-7mm} 
{}+\frac{\alpha_s^4(\mu_R^2)}{\pi} \frac{B_{\rm QED}}{N_c (N_c^2-1)}
\left[\frac{1}{s_4}\right]_{+}
\left[{\rm Re}{\Gamma'}_{11}^{gg}-2{\rm Re}{\Gamma'}_{22}^{gg}
+\left(1-\frac{2t_1u_1}{s^2}\right)N_c^2{\rm Re}{\Gamma'}_{22}^{gg}
+\frac{N_c^3}{4}{\Gamma'}_{31}^{gg}\frac{(t_1^2-u_1^2)}{s^2}\right]
\nonumber \\ && \quad \times
\left[C_A\ln\left(\frac{t_1 u_1}{m^4}\right)
\ln\left(\frac{\mu_F^2}{m^2}\right)
+\frac{\beta_0}{2}\ln\left(\frac{\mu_R^2}{\mu_F^2}\right)
+\frac{\beta_0}{4}\ln\left(\frac{\mu_R^2}{m^2}\right)\right]
\nonumber \\ && \hspace{-7mm}
{}+\sigma^B_{gg\rightarrow Q{\bar Q}} 
\left(\frac{\alpha_s(\mu_R^2)}{\pi}\right)^2
\delta(s_4)\left[\left(-2\zeta_2 C_A^2
+\frac{1}{2}C_A^2 \ln^2\left(\frac{t_1 u_1}{m^4}\right)
-\frac{5}{8}\beta_0 C_A \ln\left(\frac{t_1 u_1}{m^4}\right)\right)
\ln^2\left(\frac{\mu_F^2}{m^2}\right)\right.
\nonumber \\ && \quad \quad \quad \left.
{}+\frac{3}{4}\beta_0 C_A \ln\left(\frac{t_1 u_1}{m^4}\right)
\ln\left(\frac{\mu_F^2}{m^2}\right)\ln\left(\frac{\mu_R^2}{m^2}\right)
+\frac{3}{16}\beta_0^2 \ln^2\left(\frac{\mu_R^2}{\mu_F^2}\right)\right] \, ,
\eeqa
where  ${\hat T}^{(1)\, gg}_{\overline {\rm MS}}$ is defined in analogy to its
counterpart in the $q{\bar q}$ channel.

At the Tevatron, with $\sqrt{S}=1.8$ TeV and $\mu_F=\mu_R=m=175$ GeV/c$^2$,
the exact NLO cross section for the $gg$ channel is 0.55 pb and the NNLO-NNLL 
corrections  provide an additional 0.32 pb. 
The relative size of the NNLO corrections for the $gg$ channel compared to NLO
is much greater than for the $q \bar q$ channel. This is because of the 
different color coefficients in the expressions for the two channels, as
is obvious from the coefficients of the leading logarithms.

NNLO results for the $gg$ channel in PIM kinematics are presented 
in Appendix C (see also Ref. \cite{Kidonakis:2000ze}).

\subsection{Top production at the Tevatron}

In this subsection we add the numerical contributions from the $q{\bar q}$
and $gg$ partonic channels and
present some numerical results for the top quark total cross section 
and transverse momentum distributions at the Tevatron
(see Appendix B for a discussion of the hadronic calculation).
We use the CTEQ5M parton densities \cite{Lai:2000wy}.

\begin{figure}
\centerline{
\psfig{file=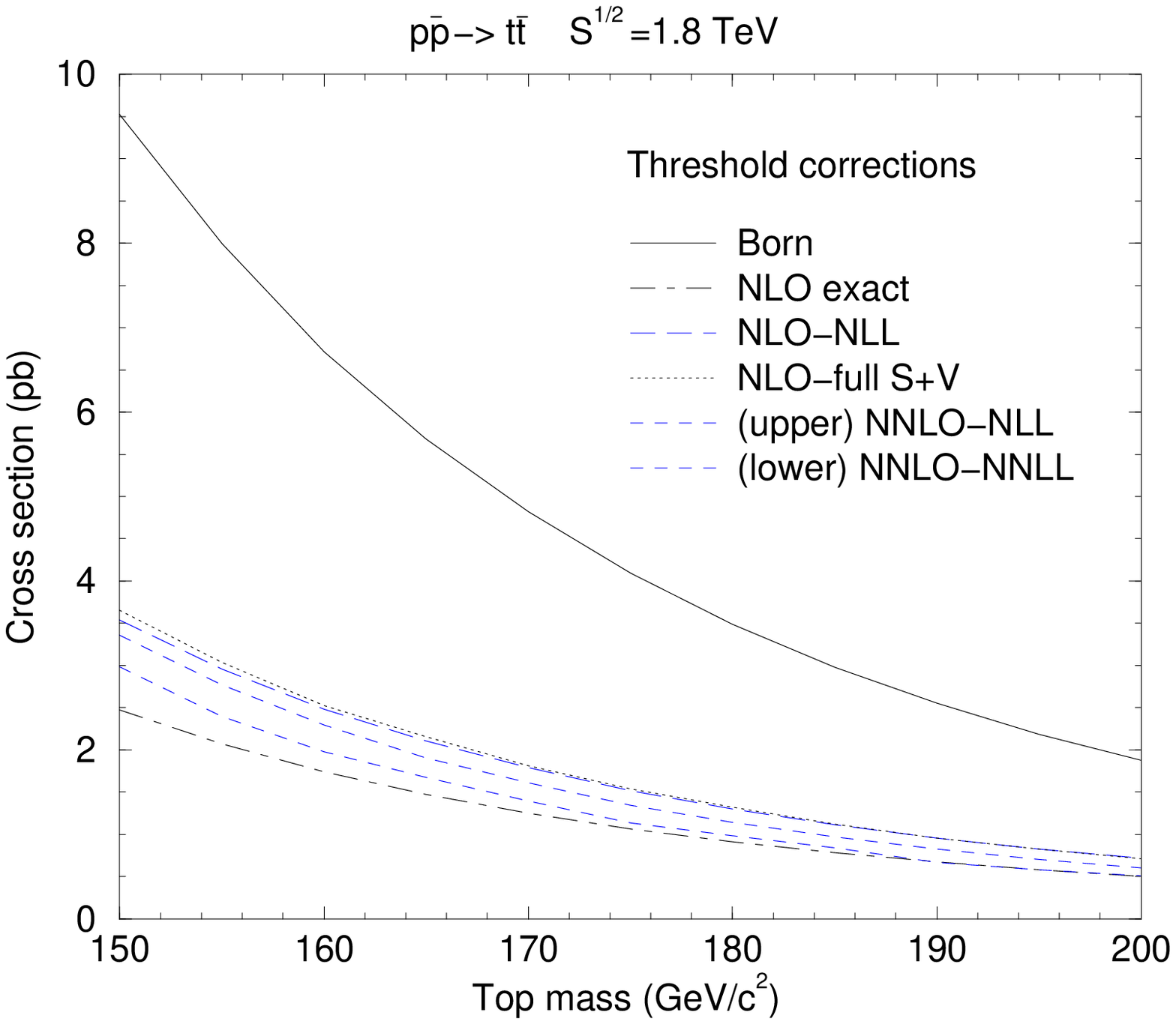,height=3.8in,width=4.5in,clip=}}
{Fig. 3. The Born, NLO, and NNLO corrections 
for top quark production at the Tevatron
with $\sqrt{S}=1.8$ TeV.} 
\label{fig3}
\end{figure}

In Fig. 3 we plot the Born term and the NLO and NNLO corrections
for top quark production at the Tevatron
with $\sqrt{S}=1.8$ TeV as a function of the top mass. At NLO we show
the exact corrections as well as the NLL threshold corrections and the 
full S+V threshold corrections. At NNLO we show results with both 
NLL and NNLL accuracy.

\begin{figure}
\centerline{
\psfig{file=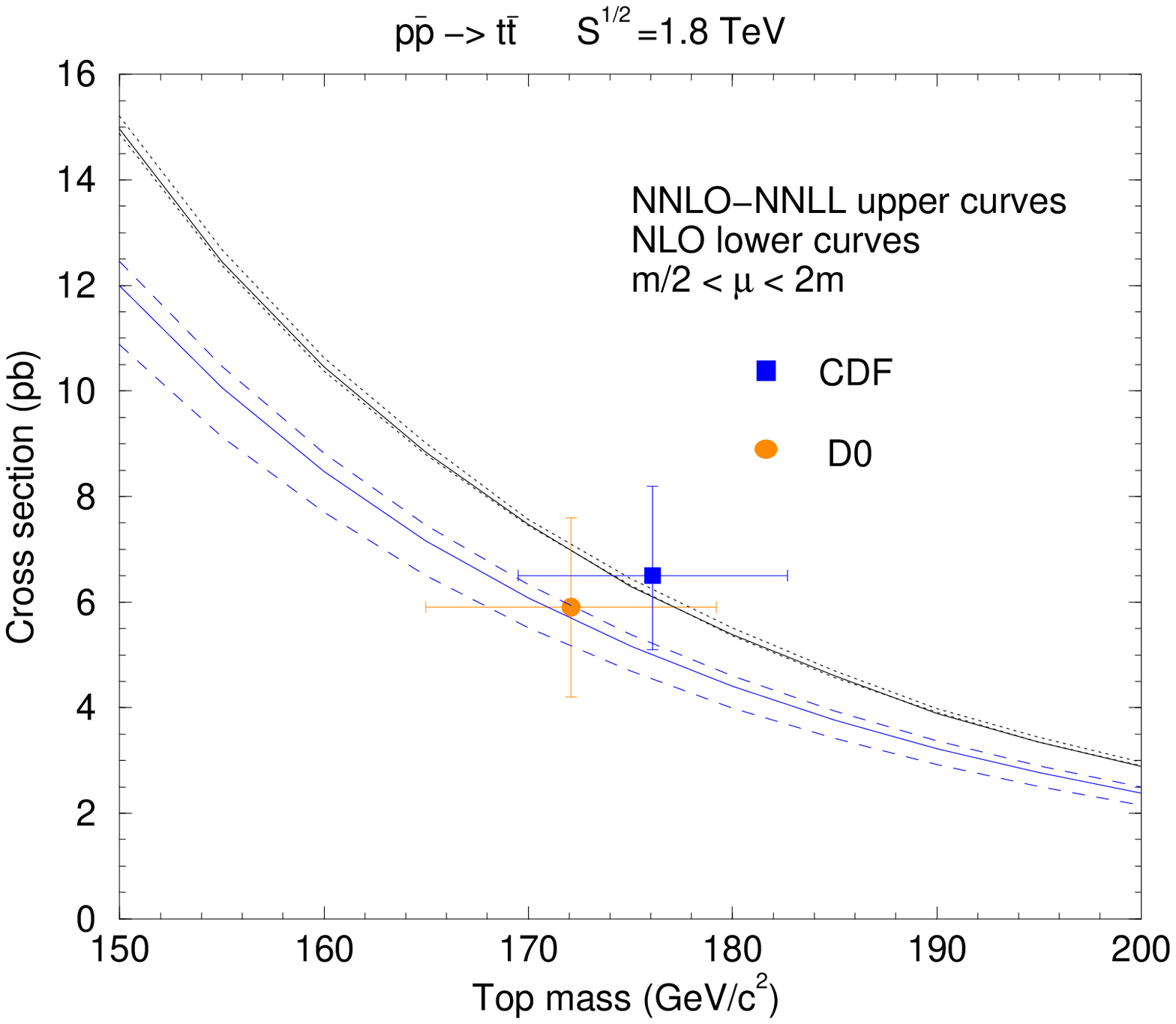,height=3.8in,width=4.5in,clip=}}
{Fig. 4. The total cross section for top quark production at the Tevatron
with $\sqrt{S}=1.8$ TeV. The labels are as in Fig. 1. 
Recent results from CDF and D0 are also shown.}
\label{fig4}
\end{figure}

In Fig. 4 we plot the exact NLO and the NNLO-NNLL ${\overline{\rm MS}}$
top quark cross section at the Tevatron
with $\sqrt{S}=1.8$ TeV as a function of the top mass. 
We note a dramatic decrease of the scale dependence 
of the cross section when we include the NNLO-NNLL corrections.
We observe that the NNLO cross section is uniformly above the 
NLO cross section for all scale choices (we stress that there is 
no field-theoretical reason for the NNLO results to lie within the 
NLO results). 
We also show recent results from CDF \cite{Blusk:2000zz} 
and D0 \cite{Abbott:1999mr} and note the agreement
between experiment and theory.
In Fig. 5 we show the corresponding results for the upgraded Tevatron
with $\sqrt{S}=2.0$ TeV.

In Table 2 we list the exact NLO and the NNLO-NNLL 
total cross sections in pb for top quark production
at the Tevatron with $\sqrt{S}=1.8$ TeV and 2.0 TeV,
a top mass $m=175$ GeV$/c^2$, and scale $\mu=m,m/2,2m$. 
The NNLO-NNLL cross section with $\sqrt{S}=1.8$ TeV
is 6.3 pb versus 5.2 pb at NLO,
an enhancement of over $20 \%$, at $\mu=m$. 
Good agreement is observed with recent results from 
CDF: $\sigma=6.5^{+1.7}_{-1.4}$ pb with $m=176.1 \pm 6.6$ GeV/$c^2$
\cite{Blusk:2000zz};  and D0: $\sigma=5.9 \pm 1.7$ pb 
with $m=172.1 \pm 7.1$ GeV/$c^2$ \cite{Abbott:1999mr}.
Similar enhancements are noted for the upgraded Tevatron energy.

\begin{table}[htb]
\begin{center}
\begin{tabular}{|c|c|c|} \hline
 $p{\bar p}\rightarrow t{\bar t}$  & $\sqrt{S}=1.8$ TeV 
& $\sqrt{S}=2.0$ TeV \\ \hline 
$\mu=\mu_F=\mu_R$   & NLO \quad NNLO & NLO  \quad NNLO \\ \hline
 $\mu=m/2$ & 5.4 \quad 6.4 & 7.4 \quad 8.9 \\ \hline
 $\mu=m$   & 5.2 \quad 6.3 & 7.1 \quad 8.8 \\ \hline
 $\mu=2m$  & 4.7 \quad 6.3 & 6.5 \quad 8.8 \\ \hline
\end{tabular}
\end{center}
\caption[]{The hadronic $t \overline t$ production cross section 
in pb for $p \overline p$ collisions with $\sqrt{S} = 1.8$ TeV and 2.0 TeV,
and $m=175$ GeV/$c^2$.}
\end{table}

\begin{figure}
\centerline{
\psfig{file=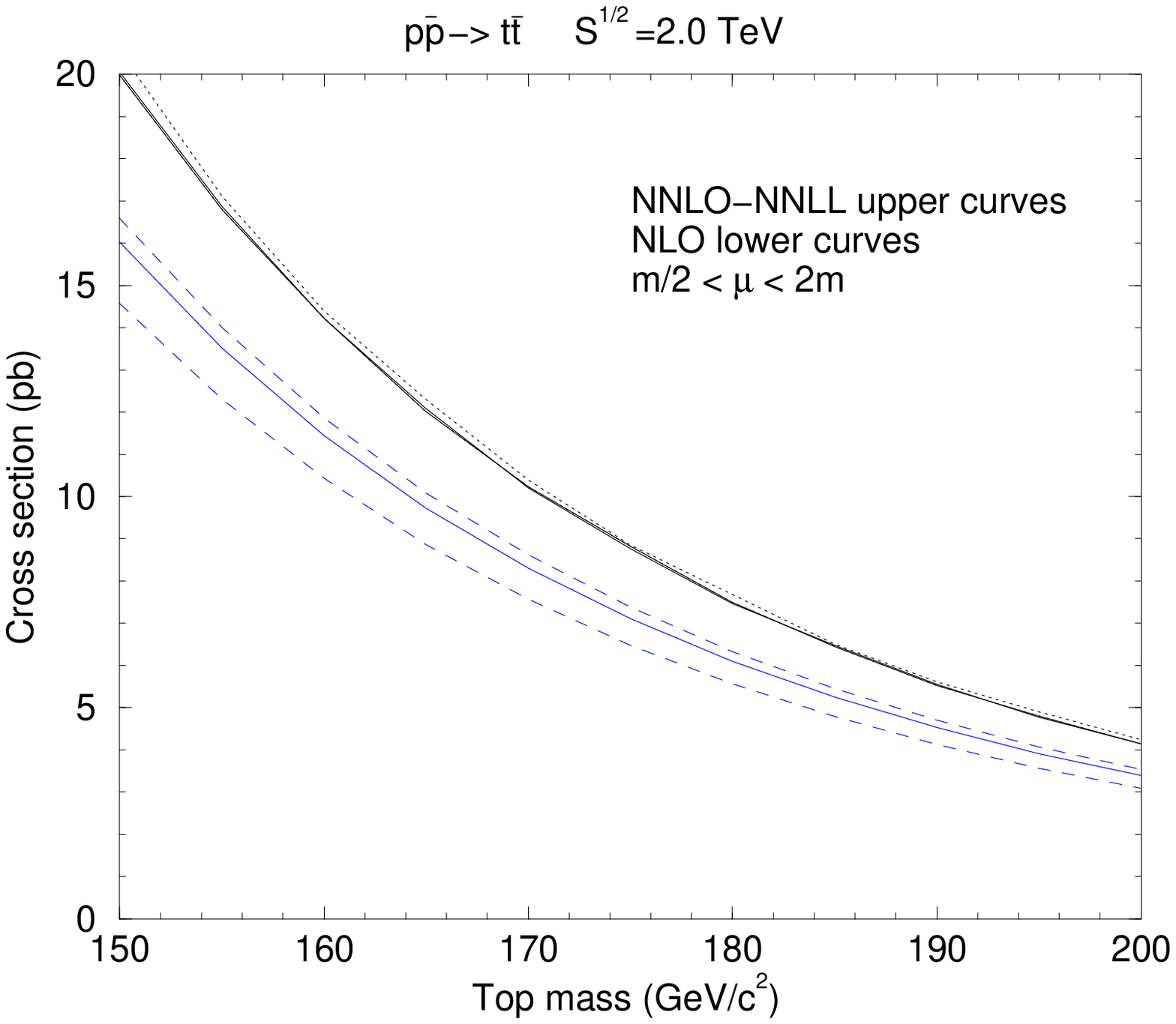,height=3.8in,width=4.5in,clip=}}
{Fig. 5.  The total cross section for top quark production at the Tevatron
with $\sqrt{S}=2.0$ TeV. The labels are as in Fig. 1.}
\label{fig5}
\end{figure}

We would like to stress that the significantly reduced scale dependence 
should not be interpreted as an equivalent reduction of the uncertainty
in the value of the cross section. Other sources of error such as 
from parton distributions, subleading logarithms, and distance from threshold, 
can provide more uncertainty than the scale variation; 
and those errors cannot all be calculated precisely at present. 
This is why we do not give more than one significant figure after 
the decimal point in the numbers of Table 2.
However, it is gratifying to see that perturbation
theory behaves as we would expect it to \cite{Oderda:1999im}: 
at higher orders the scale variation decreases.  
Since the effect of subleading logarithms is the greatest
uncertainty in the calculation, the total $t{\bar t}$ cross section
can be written with a minimal error estimate as 
$6.3^{+0.1}_{-0.4}$ pb  at $\sqrt{S}=1.8$ TeV 
and as $8.8^{+0.1}_{-0.5}$ pb at $\sqrt{S}=2.0$ TeV, 
where the larger lower error indicates the uncertainty from subleading terms.

Our formalism allows the calculation of any relevant  differential 
cross section. Transverse momentum and rapidity distributions
with leading logarithmic resummation have been presented for top
production at the Tevatron in Ref. \cite{Kidonakis:1995wz}.
The exact NLO and the NNLO-NNLL top quark transverse 
momentum ($p_T=t_1u_1/s^2-m^2$) 
distributions at the Tevatron, with $\sqrt{S}=1.8$ TeV and $2.0$ TeV, 
and $m=175$ GeV/$c^2$ are shown in Fig. 6, again in the 
${\overline {\rm MS}}$ scheme.
We note an overall enhancement at NNLO with little change of shape.
Similar conclusions are also reached for the rapidity distributions
\cite{Kidonakis:2000ph}.

\begin{figure}
\centerline{
\psfig{file=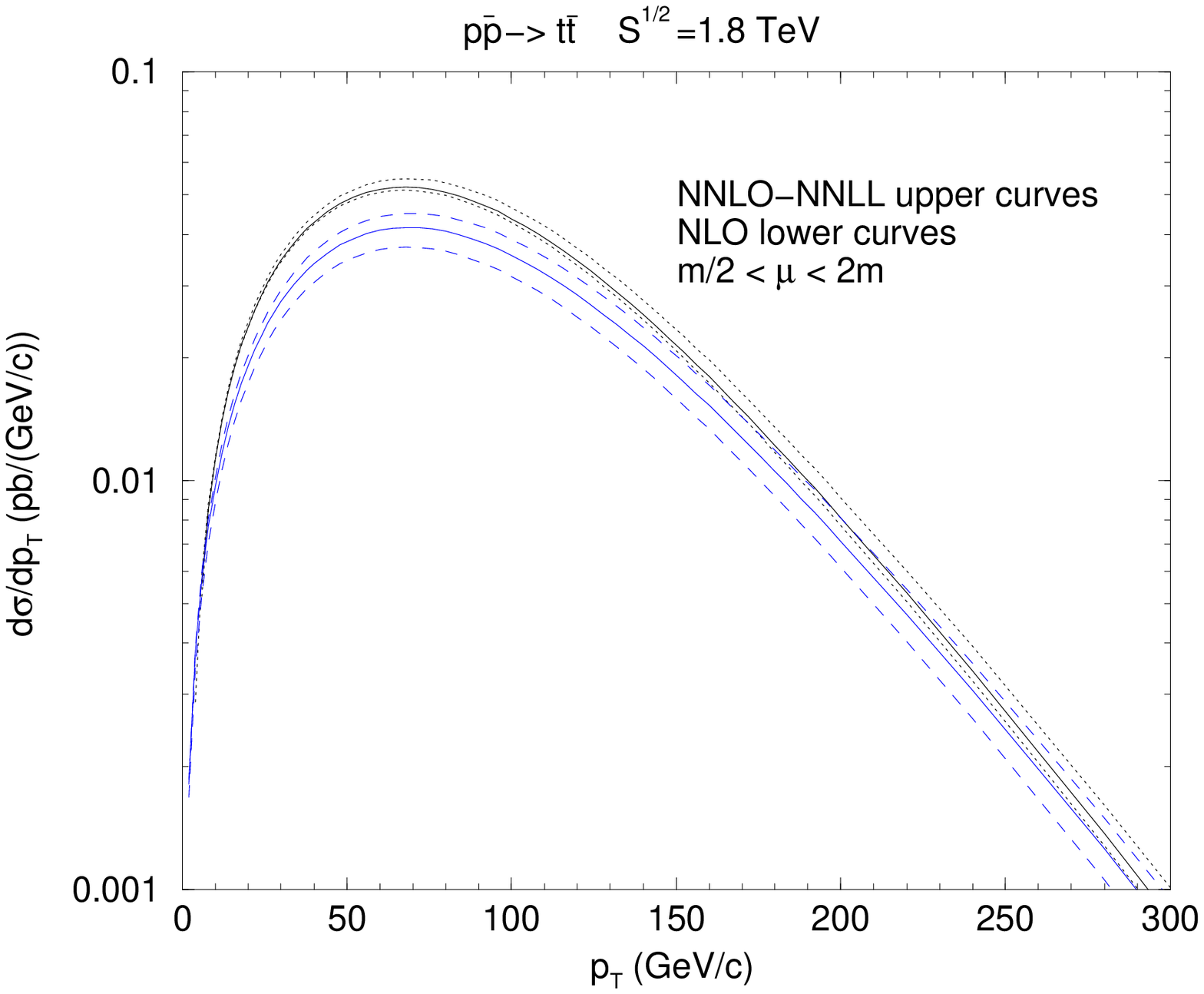,height=2.8in,width=3.2in,clip=} \hspace{5mm}
\psfig{file=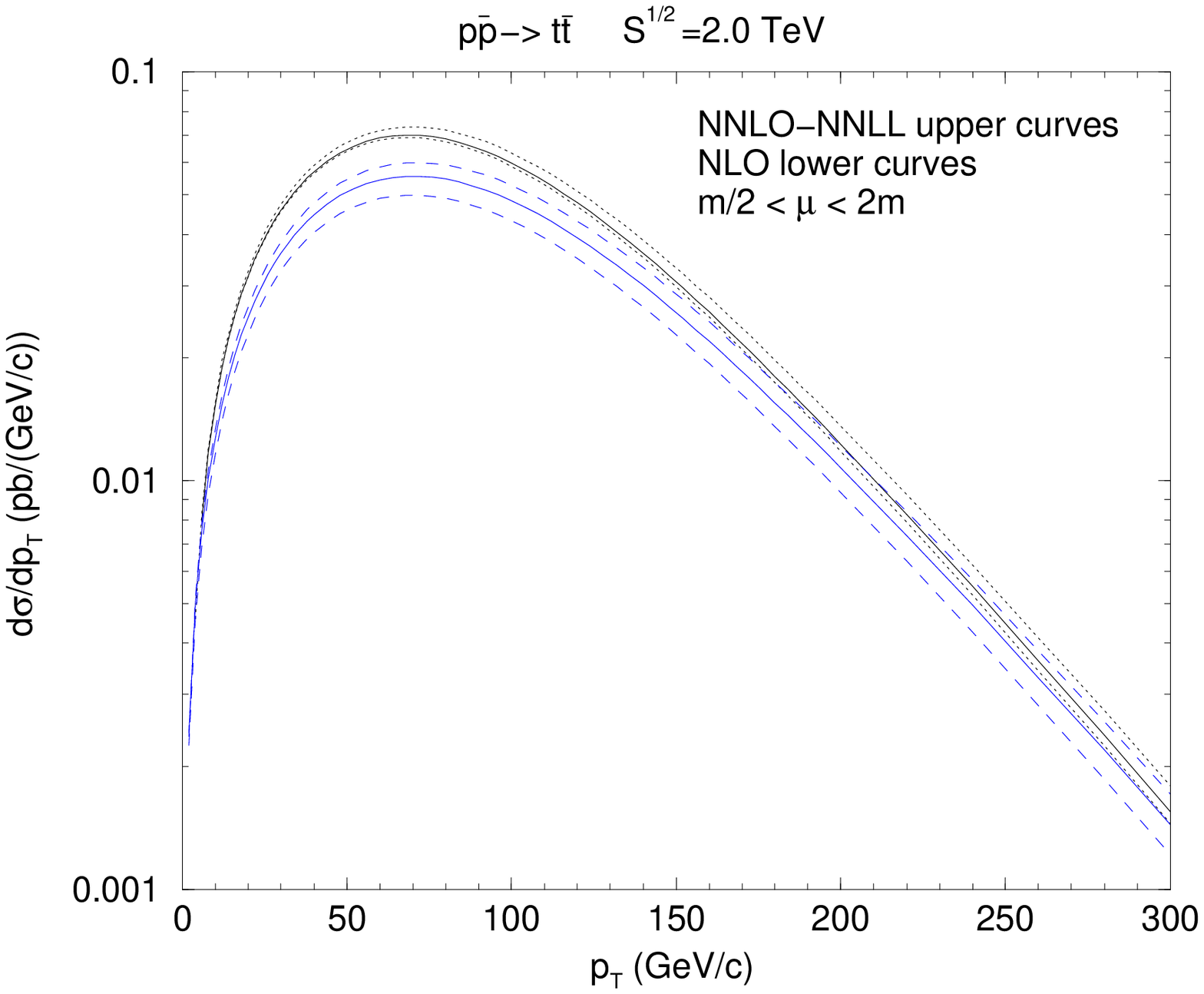,height=2.8in,width=3.2in,clip=}}
{Fig. 6.  Top quark transverse momentum distribution
at the Tevatron, with $\sqrt{S}=1.8$ TeV and $2.0$ TeV,
and $m=175$ GeV/$c^2$. The labels are as in Fig. 1.}
\label{fig6}
\end{figure}

Finally, we note that threshold resummation is also relevant for bottom
quark production at the HERA-B experiment. Leading logarithmic resummed
results for the bottom quark total cross section and differential distributions
have been presented in Refs. \cite{Kidonakis:1995uu,Kidonakis:1996jm};
for the NLL resummed cross section 
see Refs. \cite{Kidonakis:1997zd,Bonciani:1998vc}.
At NNLO-NNLL with $\sqrt{S}=41.6$ GeV and $\mu=m=4.75$ GeV$/c^2$ 
we find a cross section for that experiment of 42 nb, while the NLO
cross section is 18 nb. 

\mysection{N$^3$LO threshold corrections}

We now go beyond the NNLO corrections and expand the resummed cross section
to next-to-next-to-next-to-leading order at NNLL accuracy.

For the $q{\bar q}$ channel in the $\overline{\rm MS}$ scheme 
the N$^3$LO-NNLL threshold corrections are given by
\beqa
{\hat \sigma}^{\overline{\rm MS} \, (3)}_{q{\bar q}\rightarrow Q{\bar Q}}
(s_4,m^2,s,t_1,u_1,\mu_F,\mu_R)&=&\sigma^B_{q{\bar q}\rightarrow Q{\bar Q}}
\left(\frac{\alpha_s(\mu_R^2)}{\pi}\right)^3
\left\{8C_F^3\left[\frac{\ln^5(s_4/m^2)}{s_4}\right]_{+} \right.
\nonumber \\ && \hspace{-55mm}
{}+20 C_F^2\left[{\rm Re} {\Gamma'}_{22}^{q{\bar q}}-C_F
+C_F\ln\left(\frac{sm^2}{t_1u_1}\right)
-C_F\ln\left(\frac{\mu_F^2}{m^2}\right)-\frac{\beta_0}{6}\right]
\left[\frac{\ln^4(s_4/m^2)}{s_4}\right]_{+}
\nonumber \\ && \hspace{-55mm}
{}+\left[8C_F^2 c_1+4C_F c_2^2+16 C_F {\Gamma'}_{12}^{q{\bar q}} 
{\Gamma'}_{21}^{q{\bar q}}
+\frac{\beta_0^2}{3}C_F-64C_F^3 \zeta_2 +8 C_F^2 K \right.
\nonumber \\ && \hspace{-55mm} \left. \left.
+4 \beta_0 C_F \left(-\frac{c_2}{3}+C_F\ln\left(\frac{t_1u_1}{sm^2}\right)
+C_F\ln\left(\frac{\mu_R^2}{m^2}\right)+C_F-{\rm Re} 
{\Gamma'}_{22}^{q{\bar q}}\right)\right] 
\left[\frac{\ln^3(s_4/m^2)}{s_4}\right]_{+}\right\}
\nonumber \\
+{}{\cal O}\left(\left[\frac{\ln^2(s_4)/m^2)}{s_4}\right]_+\right)\, ,
\label{NNNLOqq}
\eeqa
where $c_1,c_2$ are defined below Eq. (\ref{NLOb}).
Note that at NNLL accuracy there are no cubic terms in $\Gamma_S$;
they start contributing at ${\cal O}([(\ln^2(s_4)/m^2)/s_4]_+)$. 
One can also derive terms involving the factorization and renormalization 
scales at lower powers of the logarithms as was explained in Section 3.

For the $gg$ channel the leading logarithms at N$^3$LO are of course
the same as for the $q{\bar q}$ channel with the substitution
$C_F \rightarrow C_A$ in the coefficients. Beyond leading logarithms
the more complex color structure of the hard scattering for the $gg$ channel
makes the calculation more lengthy, as is evident already
at NNLO,  and will not be pursued here.

In the DIS scheme the corresponding result is 
\beqa
{\hat \sigma}^{{\rm DIS} \, (3)}_{q{\bar q}\rightarrow Q{\bar Q}}
(s_4,m^2,s,t_1,u_1,\mu_F,\mu_R)&=&\sigma^B_{q{\bar q}\rightarrow Q{\bar Q}}
\left(\frac{\alpha_s(\mu_R^2)}{\pi}\right)^3
\left\{C_F^3\left[\frac{\ln^5(s_4/m^2)}{s_4}\right]_{+} \right.
\nonumber \\ && \hspace{-55mm}
{}+5 C_F^2\left[{\rm Re} {\Gamma'}_{22}^{q{\bar q}}-\frac{C_F}{4}
+\frac{1}{2}C_F\ln\left(\frac{s^2}{t_1u_1}\right)
-C_F\ln\left(\frac{\mu_F^2}{m^2}\right)-\frac{\beta_0}{4}\right]
\left[\frac{\ln^4(s_4/m^2)}{s_4}\right]_{+}
\nonumber \\ && \hspace{-55mm}
{}+\left[2C_F^2 c_1'+2C_F {c_2'}^2+8 C_F {\Gamma'}_{12}^{q{\bar q}} 
{\Gamma'}_{21}^{q{\bar q}}
+\frac{7}{24}\beta_0^2 C_F-8C_F^3 \zeta_2 +2 C_F^2 K \right.
\nonumber \\ && \hspace{-55mm} \left. \left.
+\beta_0 C_F \left(-c_2'+\frac{3}{2}C_F\ln\left(\frac{t_1u_1}{m^4}\right)
+C_F\ln\left(\frac{\mu_R^2}{s}\right)+\frac{5}{4}C_F-2{\rm Re} 
{\Gamma'}_{22}^{q{\bar q}}\right)\right] 
\left[\frac{\ln^3(s_4/m^2)}{s_4}\right]_{+}\right\}
\nonumber \\
+{}{\cal O}\left(\left[\frac{\ln^2(s_4)/m^2)}{s_4}\right]_+\right)\, ,
\eeqa
where $c_1'=c^{(1) \, q {\bar q} \, {\rm S+V}}_{\rm DIS}$, and
$c_2'=2 {\rm Re} {\Gamma'}_{22}^{q{\bar q}}
-C_F/2+C_F\ln(s^2/(t_1u_1))-2C_F\ln(\mu_F^2/m^2)$.

We can extend our study of subleading logarithms to N$^3$LO.
The N$^3$LO threshold ${\overline{\rm MS}}$ corrections 
for $q{\bar q} \rightarrow Q {\bar Q}$ in 
moment space are given in shorthand notation by
\beqa
{\hat\sigma}^{(3)}(N)&=&\sigma^B\frac{\alpha_s^3}{6\pi^3}
\left\{\left[c_1 -c_2 \ln {\tilde N} 
+\frac{c_3}{2} (\ln^2 {\tilde N}+\zeta_2)\right]^3 \right.
\nonumber \\ && \left.
{}+{\tilde F'}(\beta_0,\Gamma_S^2,\Gamma_S^3,K,3-{\rm loop}) 
\right\} \, ,
\label{NNNLOms}
\eeqa
i.e. by the cube of the terms in curly brackets in Eq. (\ref{NLOmom})
plus a function $\tilde F'$ that gives the $\beta_0$ terms that come from
changing the argument in the running coupling,
the two-loop $K$ terms, additional square and cubic terms in the 
soft anomalous dimension matrix elements,
and two- and three-loop $\Gamma_S$ and other terms.
Note that we have also absorbed in $\tilde F'$ the terms 
$2T_1^3-3T_1^2[c_1-c_2 \ln{\tilde N}+(c_3/2) (\ln^2 {\tilde N}+\zeta_2)]$,
with $T_1=c_1(\mu=m)$.
We then rewrite Eq. (\ref{NNNLOms}) in terms of $I_5,I_4,I_3,I_2,I_1$, 
and $I_0$ defined in Appendix A. 

Inverting back to momentum space, we have 
\beqa
{\hat\sigma}^{(3)}(s_4)&=&\sigma^B\frac{\alpha_s^3}{\pi^3}
\left\{\frac{1}{8} c_3^3 \left[\frac{\ln^5(s_4/m^2)}{s_4}\right]_{+}
+\frac{5}{8}c_3^2 c_2 \left[\frac{\ln^4(s_4/m^2)}{s_4}\right]_{+} \right.
\nonumber \\ &&
{}+\left(c_3 c_2^2+\frac{c_1c_3^2}{2}-\zeta_2c_3^3\right)
\left[\frac{\ln^3(s_4/m^2)}{s_4}\right]_{+}
\nonumber \\ &&
{}+\left(\frac{3}{2} c_1 c_2 c_3-3\zeta_2 c_3^2 c_2+\frac{c_2^3}{2}
+\frac{5}{2}c_3^3 \zeta_3\right)
\left[\frac{\ln^2(s_4/m^2)}{s_4}\right]_{+}
\nonumber \\ &&
{}+\left[\frac{c_1^2c_3}{2}+c_1c_2^2-\zeta_2c_3^2c_1
-\frac{5}{2}\zeta_2 c_3 c_2^2+5 \zeta_3 c_3^2 c_2
+\frac{5}{4} \zeta_2^2 c_3^3 -\frac{15}{4}c_3^3 \zeta_4\right]
\left[\frac{\ln(s_4/m^2)}{s_4}\right]_{+}
\nonumber \\ && 
{}+\left[\frac{1}{2}c_2 c_1^2+3 c_3^3 \zeta_5
-\frac{15}{4} c_3^2 c_2 \zeta_4 -2 c_3^3 \zeta_2 \zeta_3
+c_1 c_3^2 \zeta_3 + 2c_3 c_2^2 \zeta_3 \right.
\nonumber \\ && \left.
{}+\frac{5}{4} c_3^2 c_2 \zeta_2^2-c_1 c_2 c_3 \zeta_2
-\frac{1}{2} \zeta_2 c_2^3\right] \left[\frac{1}{s_4}\right]_+
\nonumber \\ &&
{}+\left[\frac{c_1^3}{6}-\frac{5}{2}c_3^3 \zeta_6
+3 c_3^2 c_2 \zeta_5+\frac{3}{2} c_3^3 \zeta_2 \zeta_4
-\frac{3}{2} c_3 c_2^2 \zeta_4-\frac{3}{4} c_3^2 c_1 \zeta_4 \right.
\nonumber \\ &&
{}+\frac{5}{6} \zeta_3^2 c_3^3-2 \zeta_2 \zeta_3 c_2 c_3^2
+c_1 c_2 c_3 \zeta_3+\frac{c_2^3}{3} \zeta_3 -\frac{1}{6}c_3^3\zeta_2^3
\nonumber \\ && \left. 
{}+\frac{1}{4}c_1 c_3^2 \zeta_2^2+\frac{1}{2}c_3 c_2^2 \zeta_2^2
-\frac{1}{2} c_1 c_2^2 \zeta_2\right] \delta(s_4) 
\nonumber \\ && \left.
{}+F'(\beta_0,\Gamma_S^2,\Gamma_S^3,K,3-{\rm loop}) \right\} \, ,
\label{3l}
\eeqa
where $F'$ comes from the inversion of ${\tilde F'}$ and starts
contributing at NLL and higher accuracy.
One can easily see that the above equation, with the appropriate 
explicit expression for $F'$, agrees with the N$^3$LO-NNLL expansion
given in Eq. (\ref{NNNLOqq}). Of course we cannot derive all of the
$[(\ln^2(s_4/m^2))/s_4]_+$ and lower terms in Eq. (\ref{3l}) because
of unknown 3-loop corrections in $F'$.

Again, let us see what happens if one keeps the logarithms only at a certain
accuracy.
At NLL accuracy for ${\hat \sigma}^{(3)}$ 
the subleading terms from the inversion
(keeping only $\zeta$ terms and no $\beta_0,K,\gamma_E$ terms as discussed
in Section 3.3) are:
\beqa
&& \sigma^B \frac{\alpha_s^3}{\pi^3} \left\{-\frac{5}{4}\zeta_2 c_3^3
\left[\frac{\ln^3(s_4/m^2)}{s_4}\right]_{+}
+\left(\frac{5}{2} \zeta_3 c_3^3 -\frac{15}{4} c_3^2 c_2 \zeta_2 \right)
\left[\frac{\ln^2(s_4/m^2)}{s_4}\right]_{+} \right.
\nonumber \\ && \hspace{-10mm}
{}+\left[\frac{15}{8}(\zeta_2^2-2\zeta_4) c_3^3+5 c_3^2 c_2 \zeta_3\right]
\left[\frac{\ln(s_4/m^2)}{s_4}\right]_{+}
+\left[\left(-\frac{5}{2}\zeta_2 \zeta_3 +3 \zeta_5\right) c_3^3
+\frac{15}{8} c_3^2 c_2 (\zeta_2^2-2 \zeta_4)\right]
\left[\frac{1}{s_4}\right]_{+}
\nonumber \\ && \left.
{}+\left[\frac{5}{2}c_3^3\left(-\frac{\zeta_2^3}{8}+\frac{\zeta_3^2}{3}
+\frac{3}{4}\zeta_2 \zeta_4 - \zeta_6\right)
+c_3^2 c_2 \left(-\frac{5}{2}\zeta_2 \zeta_3+3 \zeta_5\right)\right]
\delta(s_4) \right\} \, .
\eeqa

At NNLL accuracy for ${\hat \sigma}^{(3)}$ the corresponding subleading terms 
from the inversion to momentum space are:
\beqa
&& \sigma^B \frac{\alpha_s^3}{\pi^3} \left\{
\left(\frac{5}{2} \zeta_3 c_3^3 -\frac{15}{4} c_3^2 c_2 \zeta_2 \right)
\left[\frac{\ln^2(s_4/m^2)}{s_4}\right]_{+} \right.
\nonumber \\ &&
{}+\left[3\left(-\frac{5}{4}\zeta_4+\frac{3}{8}\zeta_2^2\right) c_3^3
+5 c_3^2 c_2 \zeta_3 -3\zeta_2 c_3 c_2^2 -\frac{3}{2} \zeta_2 c_1 c_3^2
-48 C_F \zeta_2 {\Gamma'}_{12}^{q{\bar q}} {\Gamma'}_{21}^{q{\bar q}}\right]
\left[\frac{\ln(s_4/m^2)}{s_4}\right]_{+}
\nonumber \\ &&
{}+\left[(-2\zeta_2 \zeta_3 +3 \zeta_5) c_3^3
+\frac{15}{8} c_3^2 c_2 (\zeta_2^2-2 \zeta_4)
+2 \zeta_3 c_3 c_2^2 + \zeta_3 c_1 c_3^2
+32 C_F \zeta_3 {\Gamma'}_{12}^{q{\bar q}} {\Gamma'}_{21}^{q{\bar q}}\right]
\left[\frac{1}{s_4}\right]_{+}
\nonumber \\ && 
{}+\left[\frac{1}{2}c_3^3\left(-\frac{1}{4}\zeta_2^3+\frac{5}{3}\zeta_3^2
+3 \zeta_2 \zeta_4 - 5 \zeta_6\right)
+c_3^2 c_2 \left(-\frac{5}{2}\zeta_2 \zeta_3+3 \zeta_5\right) \right.
\nonumber \\ && \left. \left.
{}+\frac{3}{4}\left(c_3 c_2^2+\frac{c_1 c_3^2}{2}\right)
(\zeta_2^2-2\zeta_4)
+12 C_F {\Gamma'}_{12}^{q{\bar q}} {\Gamma'}_{21}^{q{\bar q}}
(\zeta_2^2-2\zeta_4)
\right] \delta(s_4) \right\} \, .
\eeqa

A comparison of both the NLL and the NNLL subleading terms
with Eq. (\ref{3l}) shows that most of these terms have incorrect coefficients.
The subleading terms bring down the value
of the ${\overline {\rm MS}}$ N$^3$LO corrections for 
$q{\bar q} \rightarrow t {\bar t}$ at the Tevatron with $\sqrt{S}=1.8$ TeV
and $m=175$ GeV/$c^2$ from 0.9 pb (1.1 pb) at NNLL (NLL) to around 0.3 pb. 
As we discussed in Section 3.3, retaining subleading terms with
incorrect coefficients in the expansions can produce misleading results. 
We also note that in the DIS scheme
the corresponding corrections (without subleading terms) are 0.2 pb, 
again smaller than the ${\overline {\rm MS}}$ result because of 
the specification of the DIS scheme.
 
Since we don't know the (potentially large) coefficients
of subleading powers (beyond NNLL) of the logarithms at N$^3$LO, 
in contrast to the NNLO calculation where all logarithms were determined, 
we cannot make firm numerical predictions at this order.
This also relates to the questions raised in Ref. \cite{Lai:1999ik}
as we discussed earlier.
A full N$^3$LO calculation may give substantially different (and smaller)
results from the N$^3$LO-NNLL calculation; but at present we cannot
calculate corrections beyond NNLL accuracy. 
Therefore, as we have stated before, 
for reliable numerical predictions we prefer to stop the expansion at NNLO.  

\mysection{N$^4$LO and higher-order threshold corrections}

Finally, we briefly discuss the corrections 
at next-to-next-to-next-to-next-to-leading and higher orders. 

For the $q{\bar q}$ channel in the $\overline{\rm MS}$ scheme at NLL accuracy,
the N$^4$LO threshold corrections are given by
\beqa
{\hat \sigma}^{\overline{\rm MS} \, (4)}_{q{\bar q}\rightarrow Q{\bar Q}}
(s_4,m^2,s,t_1,u_1,\mu_F,\mu_R)&=&\sigma^B_{q{\bar q}\rightarrow Q{\bar Q}}
\left(\frac{\alpha_s(\mu_R^2)}{\pi}\right)^4
\left\{\frac{16}{3}C_F^4\left[\frac{\ln^7(s_4/m^2)}{s_4}\right]_{+} \right.
\nonumber \\ && \hspace{-55mm} \left.
{}+\frac{56}{3} C_F^3 \left[{\rm Re} {\Gamma'}_{22}^{q{\bar q}}-C_F
+C_F\ln\left(\frac{sm^2}{t_1u_1}\right)
-C_F\ln\left(\frac{\mu_F^2}{m^2}\right)-\frac{\beta_0}{4}\right]
\left[\frac{\ln^6(s_4/m^2)}{s_4}\right]_{+} \right\}
\nonumber\\
+{}{\cal O}\left(\left[\frac{\ln^5(s_4)/m^2)}{s_4}\right]_+\right)\, .
\label{4l}
\eeqa
By matching with the exact NLO cross section, as we have described before,
one can derive the full NNLL terms as well. We note that no cubic 
or quartic terms in $\Gamma_S$ appear at NNLL accuracy.
A full determination of the cross section at this order would require
four-loop calculations.
The leading logarithms for the $gg$ channel at N$^4$LO again follow
from Eq. (\ref{4l}) with the substitution $C_F \rightarrow C_A$.

In the DIS scheme the corresponding NLL result is
\beqa
{\hat \sigma}^{{\rm DIS} \, (4)}_{q{\bar q}\rightarrow Q{\bar Q}}
(s_4,m^2,s,t_1,u_1,\mu_F,\mu_R)&=&\sigma^B_{q{\bar q}\rightarrow Q{\bar Q}}
\left(\frac{\alpha_s(\mu_R^2)}{\pi}\right)^4
\left\{\frac{1}{3}C_F^4\left[\frac{\ln^7(s_4/m^2)}{s_4}\right]_{+} \right.
\nonumber \\ && \hspace{-55mm} \left.
{}+\frac{7}{3} C_F^3 \left[{\rm Re} {\Gamma'}_{22}^{q{\bar q}}-\frac{C_F}{4}
+\frac{1}{2}C_F\ln\left(\frac{s^2}{t_1u_1}\right)
-C_F\ln\left(\frac{\mu_F^2}{m^2}\right)-\frac{15}{56}\beta_0\right]
\left[\frac{\ln^6(s_4/m^2)}{s_4}\right]_{+} \right\}
\nonumber\\
+{}{\cal O}\left(\left[\frac{\ln^5(s_4)/m^2)}{s_4}\right]_+\right)\, .
\eeqa
Again, numerically the corrections in the DIS scheme are smaller
than in the ${\overline{\rm MS}}$ scheme.  

The finite-order expansion procedure can be extended to arbitrarily 
high orders at NNLL accuracy. 
In general, at $n$th order in $\alpha_s$ (beyond the Born term)
the leading logarithms in the ${\overline{\rm MS}}$ scheme are 
\beq
{\hat \sigma}^{\overline{\rm MS} \, (n)}_{q{\bar q}\rightarrow Q{\bar Q}}
(s_4,m^2,s,t_1,u_1,\mu_F,\mu_R)=\sigma^B_{q{\bar q}\rightarrow Q{\bar Q}}
\left(\frac{\alpha_s(\mu_R^2)}{\pi}\right)^n
\frac{2n}{n!}(2C_F)^n\left[\frac{\ln^{2n-1}(s_4/m^2)}{s_4}\right]_{+} + ...
\label{nl}
\eeq
For the DIS scheme we only need to replace $(2C_F)^n$ by $C_F^n$ in the above
equation. For the $gg$ channel we simply replace $C_F$ by $C_A$ as discussed
before. 
It is easy to check that Eq. (\ref{nl}) reproduces the leading logarithms
in all the expansions presented in this paper.

\mysection{Conclusions}

Threshold resummation can make powerful improvements to NLO
calculations of heavy quark production cross sections. 
The analytical form of the threshold 
corrections to the fully differential cross section can be derived at NNLL
accuracy at arbitrarily high order and explicit results have been
provided in this paper through N$^4$LO. The role of subleading logarithms has
been studied in detail and it has been shown that care must be taken
to arrive at reliable numerical predictions for the cross section.  
For top quark production
at the Tevatron NNLO-NNLL predictions have been made for the 
total cross section and transverse momentum distributions. The
NNLO-NNLL corrections are significant and they dramatically reduce
the dependence of the cross section on the scale relative to NLO.

\mysection*{Acknowledgements}

I wish to thank George Sterman for drawing my attention to the 
importance of subleading logarithms in resummation prescriptions. 
I would also like to thank Eric Laenen, Jeff Owens, Jack Smith,
and Ramona Vogt for many discussions. Some of the results in 
sections 3.2 and 3.4 were derived within the collaboration of 
Ref. \cite{KLMV} and will also be discussed along with other results 
in that paper.
This work was supported in part by the U.S. Department of Energy.

\appendix
\mysection{Mellin transforms}

Here we present some useful formulas for the Mellin transforms that are
used in the resummed cross section and the finite-order expansions.

We define
\beq
I_n(N)=\int_0^1 dz \, z^{N-1} \left[\frac{\ln^n(1-z)}{1-z}\right]_+ \, .
\eeq
One may also define
\beq
I_n(N)=\int_0^\infty ds_4 \, e^{-Ns_4/m^2} 
\left[\frac{\ln^n(s_4/m^2)}{s_4}\right]_+ \, .
\eeq
The results below are identical for either definition.

Then, we have \cite{Catani:1989ne}
\beq
I_n(N)|_{N \rightarrow +\infty}=\lim_{\epsilon\to 0^+}
\left(\frac{\partial}{\partial \epsilon}\right)^n
\left[\frac{1}{\epsilon}\left(e^{\epsilon \alpha(\epsilon)}-1\right)\right]
\left[1+O\left(\frac{1}{N}\right)\right]
\eeq
where
\beq
\alpha(\epsilon)=-\ln{\tilde N}+\sum_{n=2}^{\infty}(-1)^n
\frac{\epsilon^{n-1}}{n} \zeta_n= \sum_{n=0}^{\infty} a_n \epsilon^n
\eeq 
with
\beq
a_0=-\ln{\tilde N}, \quad a_i=\frac{(-1)^{i+1}}{i+1}\zeta_{i+1}, 
\quad i=1,...,\infty \, .
\eeq
Here ${\tilde N}=N {\rm e}^{\gamma_E}$ with $\gamma_E$ the Euler constant,
$\gamma_E$=0.577...
Also $\zeta_2=\pi^2/6$, $\zeta_4=\pi^4/90$, $\zeta_6=\pi^6/945$, 
$\zeta_8=\pi^8/9450$, etc.,
while $\zeta_3=1.2020569...$, $\zeta_5=1.0369278...$, $\zeta_7=1.0083493...$, 
etc.

Then
\beqa
I_n(N)&=&\frac{a_0^{n+1}}{n+1}+n a_0^{n-1} a_1+n (n-1) a_0^{n-2} a_2
+n(n-1)(n-2)a_0^{n-3}\left(\frac{a_1^2}{2}+a_3\right)
\nonumber \\ &&
{}+n(n-1)(n-2)(n-3)a_0^{n-4}(a_1a_2+a_4)
\nonumber \\ &&
{}+n(n-1)(n-2)(n-3)(n-4)a_0^{n-5}\left(\frac{a_1^3}{6}+\frac{a_2^2}{2}
+a_1a_3+a_5\right)+...
\eeqa

The expressions for  $I_n$ have been presented up to $n=3$ in
\cite{Catani:1989ne}. Here we extend this table to $n=7$, useful
through N$^4$LO expansions:
\beqa
I_0(N)&=& -\ln{\tilde N}
\nonumber \\
I_1(N)&=&\frac{1}{2} \ln^2{\tilde N}+\frac{\zeta_2}{2}
\nonumber \\
I_2(N)&=&-\frac{1}{3} \ln^3{\tilde N}-\zeta_2 \ln{\tilde N}
-\frac{2}{3}\zeta_3
\nonumber \\
I_3(N)&=& \frac{1}{4} \ln^4{\tilde N}+\frac{3}{2} \zeta_2 \ln^2{\tilde N}
+2 \zeta_3 \ln{\tilde N} +\frac{3}{2} \zeta_4 
+\frac{3}{4} \zeta_2^2
\nonumber \\
I_4(N)&=& -\frac{1}{5} \ln^5{\tilde N} -2 \zeta_2 \ln^3{\tilde N}
-4 \zeta_3 \ln^2{\tilde N} -3 \ln{\tilde N} (\zeta_2^2+2 \zeta_4)
-4\left(\zeta_2 \zeta_3+\frac{6}{5} \zeta_5\right)  
\nonumber \\
I_5(N)&=& \frac{1}{6} \ln^6{\tilde N}+\frac{5}{2} \zeta_2 \ln^4{\tilde N}
+\frac{20}{3} \zeta_3 \ln^3{\tilde N}
+\frac{15}{2}(\zeta_2^2+2\zeta_4) \ln^2{\tilde N}
\nonumber \\ &&
{}+4(5\zeta_2 \zeta_3+6\zeta_5) \ln{\tilde N}
+5\left(\frac{\zeta_2^3}{2}+\frac{4}{3}\zeta_3^2+3\zeta_2\zeta_4
+4\zeta_6\right)
\nonumber \\
I_6(N)&=&-\frac{1}{7} \ln^7{\tilde N}-3\zeta_2 \ln^5{\tilde N}
-10 \zeta_3 \ln^4{\tilde N}-15(\zeta_2^2+2\zeta_4) \ln^3{\tilde N}
-12(6 \zeta_5+5 \zeta_2 \zeta_3) \ln^2{\tilde N}
\nonumber \\ &&
{}-15\left(8\zeta_6+\zeta_2^3+\frac{8}{3}\zeta_3^2+6\zeta_2\zeta_4\right)
\ln{\tilde N}-60 \zeta_3 \zeta_4 -\frac{720}{7} \zeta_7 -72 \zeta_2 \zeta_5
-30 \zeta_2^2 \zeta_3
\nonumber \\
I_7(N)&=&\frac{1}{8} \ln^8{\tilde N}+\frac{7}{2}\zeta_2 \ln^6{\tilde N}
+14 \zeta_3 \ln^5{\tilde N} +\frac{105}{4} (\zeta_2^2+2 \zeta_4) 
\ln^4{\tilde N}+4(42 \zeta_5+35 \zeta_2 \zeta_3) \ln^3{\tilde N}
\nonumber \\ && \hspace{-20mm}
{}+\frac{105}{2}\left(\zeta_2^3+8 \zeta_6 +\frac{8}{3} \zeta_3^2 
+6 \zeta_2 \zeta_4\right) \ln^2{\tilde N}
+4\left(126 \zeta_2 \zeta_5+105 \zeta_3 \zeta_4+180 \zeta_7
+\frac{105}{2} \zeta_2^2 \zeta_3\right) \ln{\tilde N}
\nonumber \\ && 
{}+\frac{105}{8} \zeta_2^4+336 \zeta_3 \zeta_5
+\frac{315}{2} \zeta_4 (\zeta_2^2+\zeta_4)
+420 \zeta_2 \zeta_6 +630 \zeta_8 +140 \zeta_2 \zeta_3^2 \, .
\eeqa

\mysection{Hadronic cross sections}

The double differential hadronic cross section
$d^2\sigma_{h_1h_2}/dT_1\,dU_1$ is written as a convolution
of parton distributions with the partonic differential cross section: 
\beqa
S^2\, \frac{d^2\sigma_{h_1h_2}(S,T_1,U_1)}{dT_1\,dU_1}
&=& \sum_{i,j}
\int\limits_{x_1^-}^{1}\frac{dx_1}{x_1}
\int\limits_{x_2^-}^{1}\frac{dx_2}{x_2} \,
\phi_{i/h_1}(x_1,\mu_F^2)\, \phi_{j/h_2}(x_2,\mu_F^2) \, 
s^2\, \frac{d^2{\hat{\sigma}}_{ij}(s,t_1,u_1)}
{dt_1\,du_1} \, ,
\nonumber \\
\eeqa
where the sum is over all massless parton flavors and
$\phi_{i}(x,\mu_F^2)$ are the parton distribution functions
for flavor $i$ in hadron $h$ at factorization scale $\mu_F$. 
The hadronic invariants $S,T_1,U_1$ are defined in analogy to their 
partonic counterparts.
The lower limits of the momentum fractions of the partons in the hadrons
are given by
$x_1^- = -U_1/(S+T_1)$ and $x_2^- = -x_1 T_1/(x_1 S+U_1)$.

By making a transformation from the variables $(T_1, U_1, x_1,x_2)$
to the variables $(Y,p_T^2,x_1,s_4)$, with $Y$ and $p_T$ the rapidity
and transverse momentum, via
\beq
T_1={\sqrt{S}} (p_T^2+m^2) e^{-Y} \, , \quad 
U_1={\sqrt{S}} (p_T^2+m^2) e^Y \, , \quad
x_2=\frac{s_4-x_1 T_1}{x_1 S+U_1} \, ,
\eeq
we may write the differential cross section in $p_T$ and $Y$
as 
\beq
\frac{d^2\sigma_{h_1h_2}}{dp_T^2 \, dY}=\sum_{ij} \,\frac{1}{S}
\int_{x_1^-}^1 \frac{dx_1}{x_1}
\int_0^{x_1(S+T_1)+U_1} \frac{ds_4}{s_4-x_1 T_1} \, 
\phi(x_1)  \, \phi\left(\frac{s_4-x_1 T_1}{x_1S+U_1}\right) \,
s^2 \frac{d^2{\hat \sigma}_{ij}}{dt_1 du_1} \, .
\label{dpty}
\eeq

Now, let us write the $k$th-order partonic threshold corrections
in the shorthand notation 
\beq
s^2\frac{d^2{\hat \sigma}^{(k)}_{ij}(s,t_1,u_1)}{dt_1 \: du_1} =  
\left(\frac{\alpha_s}{\pi}\right)^k 
\left\{ A^{ij}(s,t_1,u_1) \, \delta(s_4)
+\sum_{l=0}^{2k-1} a_l^{ij}(s,t_1,u_1) 
\left[\frac{\ln^l(s_4/m^2)}{s_4}\right]_+ \right\} \, .
\label{partcrosect}
\eeq

By substituting the above expression for the partonic threshold corrections
in Eq. (\ref{dpty}), we can write the hadronic $k$th-order corrections as
\beqa
\frac{d^2\sigma^{(k)}_{h_1h_2}}{dp_T^2 \, dY}&=& \sum_{ij}
\left(\frac{\alpha_s}{\pi}\right)^k \frac{1}{S} 
\int_{x_1^-}^1 \frac{dx_1}{x_1} \phi(x_1) 
\int_0^{x_1(S+T_1)+U_1} \frac{ds_4}{s_4-x_1 T_1} \, 
\phi\left(\frac{s_4-x_1 T_1}{x_1S+U_1}\right)
\nonumber \\ && \times 
\left\{A^{ij}(s_4) \, \delta(s_4)
+\sum_{l=0}^{2k-1}
a_l^{ij}(s_4) \left[\frac{\ln^l(s_4/m^2)}{s_4}\right]_+ 
\right\} \, .
\eeqa
After some rearrangements of terms and partial integrations,
we can rewrite this as
\beqa
\frac{d^2\sigma^{(k)}_{h_1h_2}}
{dp_T^2 \, dY}&=& \sum_{ij} \, \left(\frac{\alpha_s}{\pi}\right)^k 
\frac{1}{S} \sum_{l=0}^{2k-1}
\left\{ \int_{x_1^-}^1 \frac{dx_1}{x_1} \phi(x_1)
\int_0^{x_1(S+T_1)+U_1} ds_4 \, \theta(s_4-\Delta)
\right.
\nonumber \\ && \hspace{-20mm}
\times \frac{1}{s_4} \ln^l\left(\frac{s_4}{m^2}\right) \left[a_l(s_4) 
\frac{1}{s_4-x_1 T_1} \phi\left(\frac{s_4-x_1 T_1}{x_1S+U_1}\right)
-a_l(0)
\frac{1}{(-x_1 T_1)} \phi\left(\frac{-x_1 T_1}{x_1S+U_1}\right)\right]
\nonumber \\ && \hspace{-33mm} \left. 
{}+\int_{x_1^-}^1 \frac{dx_1}{x_1} \left[\frac{1}{l+1}
\ln^{l+1}\left(\frac{x_1(S+T_1)+U_1}{m^2}\right)
a_l(0) +A(0) \right]  \frac{1}{(-x_1 T_1)}
\phi(x_1) \phi\left(\frac{-x_1 T_1}{x_1S+U_1}\right) \right\}.
\eeqa

The transverse momentum distributions are then given by
\beq
\frac{d\sigma^{(k)}_{h_1h_2}}{dp_T}= 
2 p_T \frac{d\sigma^{(k)}_{h_1h_2}}{dp_T^2}
=2 p_T \int_{Y^-}^{Y^+} dY \, 
\frac{d^2\sigma^{(k)}_{h_1h_2}}{dp_T^2 \, dY} \, ,
\label{hadropt}
\eeq
where
\beq
Y^{\pm}=\pm \frac{1}{2} \ln\left(\frac{1+\beta_T}{1-\beta_T}\right)
\label{Ylimits}
\eeq
and $\beta_T=\sqrt{1-4(p_T^2+m^2)/S}$.
The total cross section can then be retrieved by integrating 
Eq. (\ref{hadropt}) over $p_T$ with lower limit $0$ and upper
limit $\sqrt{S/4-m^2}$.

\mysection{NLO and NNLO threshold corrections in PIM kinematics}

In this Appendix we present results for the NLO and NNLO expansions
of the resummed cross section in heavy-quark-pair inclusive kinematics
(see also Refs. \cite{Kidonakis:1997gm,Kidonakis:2000ze,KLMV}). 
Here the distance from threshold is measured
in terms of the variable $z=Q^2/s$, with $Q^2$ the invariant mass
squared of the heavy quark-antiquark pair, and the corresponding ``plus'' 
distributions are of the form $[(\ln^k(1-z))/(1-z)]_+$.

\subsection{NLO threshold corrections}

In the $\overline{\rm MS}$ scheme the NLO-NLL corrections 
for the $q{\bar q}\rightarrow Q{\bar Q}$ channel are
\begin{eqnarray}
{\hat \sigma}^{\overline{\rm MS} \, (1)}_{q{\bar q}\rightarrow Q{\bar Q}}
(1-z,m^2,s,t_1,u_1,\mu_F,\mu_R)&=&\sigma^B_{q{\bar q}\rightarrow Q{\bar Q}}
\frac{\alpha_s(\mu_R^2)}{\pi}\left\{4C_F\left[\frac{\ln(1-z)}{1-z}\right]_{+}
\right.
\nonumber \\ && \hspace{-55mm}
{}+\left[\frac{1}{1-z}\right]_{+} 
\left[2 {\rm Re} {\Gamma'}_{22}^{q{\bar q}}
-2C_F-2C_F \ln\left(\frac{\mu_F^2}{s}\right)\right]
\nonumber \\ && \hspace{-55mm}
\left.
{}+\delta(1-z) \left[-\frac{3}{2}C_F\ln\left(\frac{\mu_F^2}{s}\right) 
+\frac{\beta_0}{2}\ln\left(\frac{\mu_R^2}{s}\right)\right]
\right\} \, .
\end{eqnarray}
We note that in the $\delta(1-z)$ contribution
the expansion reproduces only the scale-dependent terms.  
The rest of the $\delta(1-z)$ terms can only be obtained by matching to the 
exact NLO cross section in PIM kinematics \cite{KLMV}; 
this is exactly analogous to
what was presented in Section 3 for 1PI kinematics.

In the DIS scheme the corresponding result is
\begin{eqnarray}
{\hat \sigma}^{\rm DIS \, (1)}_{q{\bar q}\rightarrow Q{\bar Q}}
(1-z,m^2,s,t_1,u_1,\mu_F,\mu_R)&=&\sigma^B_{q{\bar q}\rightarrow Q{\bar Q}}
\frac{\alpha_s(\mu_R^2)}{\pi}\left\{2C_F\left[\frac{\ln(1-z)}{1-z}\right]_{+}
\right.
\nonumber \\ && \hspace{-55mm}
{}+\left[\frac{1}{1-z}\right]_{+} \left[2 {\rm Re} {\Gamma'}_{22}^{q{\bar q}}
-\frac{C_F}{2}-2C_F \ln\left(\frac{\mu_F^2}{s}\right)\right]
\nonumber \\ && \hspace{-55mm}\left.
{}+\delta(1-z) \left[-\frac{3}{2}C_F\ln\left(\frac{\mu_F^2}{s}\right) 
+\frac{\beta_0}{2}\ln\left(\frac{\mu_R^2}{s}\right)\right]
\right\} \, .
\end{eqnarray}

For the $gg \rightarrow Q{\bar Q}$ channel the NLO-NLL corrections
in the $\overline{\rm MS}$ scheme are
\beqa
{\hat \sigma}^{\overline {\rm MS} \, (1)}_{gg \rightarrow Q{\bar Q}}
(1-z,m^2,s,t_1,u_1,\mu_F,\mu_R)&=&
\sigma^B_{gg\rightarrow Q{\bar Q}} \frac{\alpha_s(\mu_R^2)}{\pi} 
\left\{4C_A\left[\frac{\ln(1-z)}{1-z}\right]_{+}\right.
\nonumber \\ && \hspace{-45mm} \left.
{}-2C_A \ln\left(\frac{\mu_F^2}{s}\right) 
\left[\frac{1}{1-z}\right]_{+}
+\delta(1-z) \frac{\beta_0}{2} \ln\left(\frac{\mu_R^2}{\mu_F^2}\right) 
\right\}
\nonumber \\ && \hspace{-55mm}
{}+\alpha_s^3(\mu_R^2) K_{gg} B_{QED} \left[\frac{1}{1-z}\right]_{+}
\left\{N_c(N_c^2-1)\frac{(t_1^2+u_1^2)}{s^2}
\left[\left(-C_F+\frac{C_A}{2}\right)
{\rm Re} L_{\beta}\right. \right.
\nonumber \\ && \hspace{-40mm} \left.
{}+\frac{C_A}{2}\ln\left(\frac{t_1u_1}{m^2s}\right)
-C_F\right]+\frac{(N_c^2-1)}{N_c}(C_F-C_A) {\rm Re} L_{\beta}
\nonumber \\ && \hspace{-55mm} \left.
{}-(N_c^2-1)\ln\left(\frac{t_1u_1}{m^2s}\right)
+C_F \frac{(N_c^2-1)}{N_c}
+\frac{N_c^2}{2}(N_c^2-1)
\ln\left(\frac{u_1}{t_1}\right)\frac{(t_1^2-u_1^2)}{s^2} \right\} \, .
\eeqa

\subsection{NNLO threshold corrections}

We now present the NNLO corrections in PIM kinematics.
We give explicit results at NLL accuracy.
As noted in Section 3, to reach NNLL accuracy we need to derive
matching terms in PIM kinematics \cite{KLMV}.
The only other difference between the expansions in the two different 
kinematics are in the extra terms involving $\ln(t_1u_1/m^4)$ for 
the 1PI fixed-order expansions relative to the PIM expansions. 

In the $\overline {\rm MS}$ scheme, the NNLO-NLL corrections 
for  $q{\bar q}\rightarrow Q{\bar Q}$ are
\begin{eqnarray}
{\hat \sigma}^{\overline {\rm MS} \, (2)}_{q{\bar q}\rightarrow Q{\bar Q}}
(1-z,m^2,s,t_1,u_1,\mu_F,\mu_R)&=&
\sigma^B_{q{\bar q}\rightarrow Q{\bar Q}} 
\left(\frac{\alpha_s(\mu_R^2)}{\pi}\right)^2 
\left\{8 C_F^2 \left[\frac{\ln^3(1-z)}{1-z}\right]_{+} \right.
\nonumber \\ && \hspace{-55mm}
{}+\left[\frac{\ln^2(1-z)}{1-z}\right]_{+} \left[-\beta_0 C_F 
+12C_F \left({\rm Re} {\Gamma'}_{22}^{q{\bar q}}
-C_F-C_F\ln\left(\frac{\mu_F^2}{s}\right)\right)\right]
\nonumber \\ && \hspace{-55mm}
{}+\left[\frac{\ln(1-z)}{1-z}\right]_{+} \left[ C_F
\ln\left(\frac{\mu_F^2}{s}\right)
\left(-8{\rm Re} {\Gamma'}_{22}^{q{\bar q}}
+2C_F+4C_F\ln\left(\frac{\mu_F^2}{s}\right)\right)
+3 C_F \beta_0 \ln\left(\frac{\mu_R^2}{s}\right) \right]
\nonumber \\ && \hspace{-55mm} \left.
{}+\left[\frac{1}{1-z}\right]_{+} \left[
C_F\left(3C_F+\frac{\beta_0}{4}\right)\ln^2\left(\frac{\mu_F^2}{s}\right)
-\frac{3}{2} C_F \beta_0
\ln\left(\frac{\mu_R^2}{s}\right)
\ln\left(\frac{\mu_F^2}{s}\right)\right] \right\} \, .
\end{eqnarray} 
We note that, at NLL accuracy, in the $[\ln(1-z)/(1-z)]_+$ coefficient
we derive only the terms involving the factorization and renormalization 
scales, and in the $[1/(1-z)]_+$ coefficient we can only determine quadratic 
terms in the scale logarithms. 

In the DIS scheme the corresponding corrections are
\begin{eqnarray}
{\hat \sigma}^{\rm DIS \, (2)}_{q{\bar q}\rightarrow Q{\bar Q}}
(1-z,m^2,s,t_1,u_1,\mu_F,\mu_R)&=&
\sigma^B_{q{\bar q}\rightarrow Q{\bar Q}} 
\left(\frac{\alpha_s(\mu_R^2)}{\pi}\right)^2 
\left\{2 C_F^2 \left[\frac{\ln^3(1-z)}{1-z}\right]_{+} \right.
\nonumber \\ && \hspace{-55mm}
{}+\left[\frac{\ln^2(1-z)}{1-z}\right]_{+} \left[-\frac{3\beta_0}{4}C_F 
+6C_F \left({\rm Re} {\Gamma'}_{22}^{q{\bar q}}-\frac{C_F}{4}
-C_F\ln\left(\frac{\mu_F^2}{s}\right)\right)\right]
\nonumber \\ && \hspace{-55mm}
{}+\left[\frac{\ln(1-z)}{1-z}\right]_{+} 
\left[C_F \ln\left(\frac{\mu_F^2}{s}\right)
\left(-8{\rm Re} {\Gamma'}_{22}^{q{\bar q}}
-C_F+4C_F\ln\left(\frac{\mu_F^2}{s}\right)\right)
+ \frac{3}{2} C_F \beta_0 \ln\left(\frac{\mu_R^2}{s}\right)\right]
\nonumber \\ && \hspace{-55mm} \left.
{}+\left[\frac{1}{1-z}\right]_{+} \left[
C_F\left(3C_F+\frac{\beta_0}{4}\right)
\ln^2\left(\frac{\mu_F^2}{s}\right) -\frac{3}{2} C_F \beta_0 
\ln\left(\frac{\mu_R^2}{s}\right)
\ln\left(\frac{\mu_F^2}{s}\right)\right]\right\} \, .
\end{eqnarray} 

In the $\overline{\rm MS}$ scheme for the $gg \rightarrow Q{\bar Q}$
channel the NNLO-NLL corrections are
\beqa
{\hat \sigma}^{\overline {\rm MS} \, (2)}_{gg \rightarrow Q{\bar Q}}
(1-z,m^2,s,t_1,u_1,\mu_F,\mu_R)&=&
\sigma^B_{gg\rightarrow Q{\bar Q}} \left(\frac{\alpha_s(\mu_R^2)}{\pi}\right)^2
\left\{8C_A^2\left[\frac{\ln^3(1-z)}{1-z}\right]_{+}\right.
\nonumber \\ && \hspace{-55mm}\quad \quad \left.
+\left[-\beta_0 C_A -12 C_A^2 \ln\left(\frac{\mu_F^2}{s}\right) \right]
\left[\frac{\ln^2(1-z)}{1-z}\right]_{+} \right\}
\nonumber \\ && \hspace{-55mm}
{}+\frac{\alpha_s^4(\mu_R^2)}{\pi} K_{gg} B_{\rm QED} 
\left[\frac{\ln^2(1-z)}{1-z}\right]_{+} C_A \, 3(N_c^2-1)
\left\{\frac{(t_1^2+u_1^2)}{s^2} \right.
\nonumber \\ && \hspace{-55mm} \quad \times \, 
\left[N_c^2 \ln\left(\frac{t_1u_1}{m^2s}\right)
-2N_c\left(C_F-\frac{C_A}{2}\right){\rm Re}L_{\beta}-2 N_c C_F\right]
+2\frac{C_F}{N_c}
\nonumber \\ && \hspace{-55mm} \quad \left.
{}+2\ln\left(\frac{sm^2}{t_1 \, u_1}\right)
+2\frac{1}{N_c}(C_F-C_A) \,{\rm Re} L_{\beta}
+N_c^2\frac{(t_1^2-u_1^2)}{s^2}\ln\left(\frac{u_1}{t_1}\right)\right\} 
\nonumber \\ && \hspace{-55mm} 
{}+\sigma^B_{gg\rightarrow Q{\bar Q}} 
\left(\frac{\alpha_s(\mu_R^2)}{\pi}\right)^2
\left[\frac{\ln(1-z)}{1-z}\right]_{+} 
\left\{ \ln\left(\frac{\mu_F^2}{s}\right) C_A
\left[-2\beta_0 \right. \right.
\nonumber \\ && \hspace{-25mm} \left. \left.
{}+4C_A\left(\ln\left(\frac{sm^2}{t_1u_1}\right)+1
+\ln\left(\frac{\mu_F^2}{s}\right)\right)\right]
+3 C_A \beta_0 \ln\left(\frac{\mu_R^2}{s}\right)\right\}
\nonumber \\ && \hspace{-55mm}
{}+\frac{\alpha_s^4(\mu_R^2)}{\pi} K_{gg} B_{\rm QED} 
\left[\frac{\ln(1-z)}{1-z}\right]_{+} \ln\left(\frac{\mu_F^2}{s}\right) 
C_A \, 2(N_c^2-1) \left\{N_c^2 \ln\left(\frac{u_1}{t_1}\right) 
\frac{(u_1^2-t_1^2)}{s^2}\right.
\nonumber \\ && \hspace{-55mm} \quad  
{}+2N_c \left(C_F-\frac{C_A}{2}\right) ({\rm Re} L_{\beta}+1)
\frac{(t_1^2+u_1^2)}{s^2}
+\ln\left(\frac{t_1u_1}{sm^2}\right)
\nonumber \\ && \hspace{-55mm} \quad  \left.
{}+\left(2-\frac{2C_F}{N_c}\right) ({\rm Re} L_{\beta}+1)-1 \right\}
\nonumber \\ && \hspace{-55mm}
{}+\sigma^B_{gg\rightarrow Q{\bar Q}} 
\left(\frac{\alpha_s(\mu_R^2)}{\pi}\right)^2
\left[\frac{1}{1-z}\right]_{+} \left[
\frac{5}{4} C_A \beta_0 \ln^2\left(\frac{\mu_F^2}{s}\right)
-\frac{3}{2} C_A \beta_0 
\ln\left(\frac{\mu_R^2}{s}\right)
\ln\left(\frac{\mu_F^2}{s}\right)\right] \, .
\nonumber \\
\eeqa

\end{document}